# Radiative Decay of Dark Exciton Related Emission in a Sandwiched Monolayer $WSe_2$ Revealed by Room Temperature Micro and Nano Photoluminescence

Mahfujur Rahaman[1]*, Oleksandr Selyshchev[1], Yang Pan[1], Ilya Milekhin[1], Apoorva Sharma[1], Georgeta Salvan[1,2], Sibylle Gemming[1,2], Dietrich R T Zahn[1,2]

[1] Institute of Physics, Chemnitz University of Technology, 09107 Chemnitz, Germany

[2] Center for Materials, Architectures and Integration of Nanomembranes (MAIN), Chemnitz University of Technology, 09107 Chemnitz, Germany

Email address: mahfujur.rahaman@physik.tu-chemnitz.de

## Abstract

Transition metal dichalcogenides have attracted a lot of attention in recent years due to their unique indirect to direct band gap transition from bulk to monolayer thickness. Strong confinement in the out-of-plane direction enhances the Coulomb potential between the charged particles (e-h pairs) and thus increases the exciton binding energy dramatically. The lattice inversion asymmetry in a monolayer creates two non-equivalent (but degenerate in energy) band edges protected by time reversal polarisation via pseudo-spin. However, the presence of strong spin-orbit coupling in the valence band and weak spin-splitting in the conduction band results in the lowest lying exciton in $WX_2$ (X = S, Se) being spin forbidden and optically dark. Because of their long life times, dark excitons are highly attractive for quantum optics and optoelectronic applications. To date studying dark excitonic emission is limited to cryogenic temperature or required very complex experimental configurations to observe them at room temperature. Here, we demonstrate a novel approach of radiative decay of dark exciton related emissions in $WSe_2$ monolayers studied by conventional and tip-enhanced photoluminescence (TEPL) at room temperature. Monolayer $WSe_2$ flakes were sandwiched by noble metal (Au or Ag) substrates and polydimethylsiloxane (PDMS) nano-patches providing a strong local out-of-plane dipole moment with respect to the two dimensional plane. This strong dipole moment not only enhances the dark excitons in $WSe_2$, it also produces bound excitons due to extrinsic charge defects visible at room temperature. The spatial distributions of these dark exciton related emissions were studied by TEPL with a spatial resolution < 10 nm confirming the confinement of these excitons within the PDMS nano-patches. Tip-enhanced Raman scattering (TERS) investigation reveals a direct correlation between the dark excitons and the defects in $WSe_2$. Finally, by removing the nano-patches from the top of the flakes we are able to recover the optically bright excitons in the $WSe_2$ monolayer. Our approach paves the way for deep understanding and to harness excitonic properties in low dimensional semiconductors, thus offering a platform towards quantum optoelectronics.

## Introduction

Beyond graphene, transition metal dichalcogenides (TMDCs) are currently at the heart of two-dimensional (2D) material research owing to their extraordinary fundamental physical properties. Among all TMDCs, molybdenum (Mo)- and tungsten (W)-based chalcogenides (S, Se, and Te) are the most widely studied materials. When thinned down to monolayer thickness, new properties emerge off these materials, including an indirect to direct band gap transition in the visible region situated at the $K$ point of the Brillouin zone.[1] Despite being atomically thin, TMDC monolayers can absorb up to 15 % light in the visible range due to the strong light-matter interaction and are therefore highly promising for applications in optoelectronics.[2] Moreover, the loss of the inversion centre in the monolayers offers great potential for spintronic[3] and higher harmonic generation[4] applications. The dramatically reduced dielectric screening due to the strong confinement in the out-of-plane direction combined with large electron and hole effective masses creates strongly bound excitons. The exciton binding energies are in the range of a few hundreds of meV.[5] Therefore, the optical response is strongly dominated by excitons and hence provides an ideal platform for studying excitonic or many-body physics at room temperature.[6]

One important feature of TMDC monolayers is the broken inversion symmetry coupled to spin-orbit coupling (SOC). This leads to valley-spin locking and valley (and spin) polarized optical absorptions/emissions.[7-10] SOC in TMDCs creates spin splitting in both the valence band (VB) and the conduction band (CB) with opposite spins at $K$ and $K'$ band edges.[11-14] The contribution of $d_{x^2} - d_{y^2}$ and $d_{xy}$ orbitals of transition metals generate a large splitting (of the order of few hundreds meV) in the VB showing well separated optically allowed transitions from each sub-band known as A and B excitons.[15] However, CB states stem from predominantly $d_{z^2}$ orbitals. Therefore, the second order perturbation from transition metal ($d_{xz}$ and $d_{yz}$ orbitals) and chalcogen ($p_x$ and $p_y$) atoms leads to a modest spin splitting (of the order of tens of meV).[11] The conduction band spin splitting hosts both bright (optically active) and dark (optically inactive) excitonic states for both A and B excitons; the lowest lying excitons are bright (dark) in Mo (W) based TMDC monolayers.[16-17] An exciton is optically bright (dark) when the VB and CB states have the same (opposite) spin projections in the same valley for electrons. One key advantage of dark excitons is the long life time due to non-radiative decay channels and spin-flip processes.[17-19] This distinct feature offers great potential in the field of applications like Bose-Einstein Condensation (BEC)[20] or quantum computing.[21] Lowest lying dark excitons will quench the light emission by absorbing bright states particularly at low temperatures. However, via intermediate channels dark states can also become bright, restoring many body effects.[22-24] Especially, dark states when coupled to strain-localized point defects can assist to the funnelling of excitons to these defects, leading to single photon emission in TMDC monolayers.[25-26] Additionally, when localized charged defects capture dark states can brighten the excitons in the form of charge impurity states.[23] Thus, the variety of spin, valleys, and the number of complexes that can form well-resolved optically active bound states results in rich light-matter interactions and may offer plenty of useful tools for accessing the quantized valley pseudospin information. Therefore, it is of critical importance to induce radiative emission of dark excitons in TMDC monolayers for valley and spin transport and optical manipulation.

Several approaches were demonstrated in recent low temperature photoluminescence studies to induce the radiative decay of dark excitons in TMDC monolayers.[27-31] One of the studies involved applying a strong in-plane magnetic field ($\geq$ 14 T) to tilt the electron spin direction.[27-28] This approach induces a weakly allowed in-plane optical transition via the Zeeman effect. In another attempt, the out-of-plane surface plasmon polariton was coupled to the dark excitons inducing a spin-flip, thus causing radiative emission.[32] Alternatively one can also detect the dark optical transition using an objective of high numerical aperture (NA) from the sample edges.[33] It is important to note that, the weak nature of the dark-exciton related emissions limited all the above mentioned studies to be performed at cryogenic temperatures, since otherwise the small energy difference between the bright and dark emission (< 50 meV) leads to an overwhelming thermal population into the bright exciton channel. More recently, Park *et al*. demonstrated radiative emission of dark excitons at room temperature by coupling the out-of-plane transition dipole moment to a scanning probe nano-optical antenna.[34] The nano-gap between the tip and the substrate creates a strongly confined out-of-plane optical dipole moment, which facilitates the probing of dark excitons. However, in that work a complex experimental geometry was required to detect dark excitons at room temperature.

Here, we demonstrate a novel approach to induce radiative emission from dark excitons in a TMDC monolayer at room temperature, which can be detected via conventional PL spectroscopy. As a probing material we chose $WSe_2$ for two specific reasons: first, for its high bright-dark energy splitting, and secondly, for a straightforward comparison with previous works. A scheme of the investigated system is presented in Fig. 1a. The monolayer $WSe_2$ is sandwiched between an Au (or Ag) substrate and PDMS nano-patches. The combination of metal and PDMS induces a strong out-of-plane electrostatic dipole gradient in $WSe_2$ serving as a local gate as shown in Fig. 1b. PDMS is known to have a large dipole moment along its polar Si-O bonds, which can alter the local structure or dynamics of physisorbed probe molecules[35]. This particular phenomenon inspired us to test the possibility of altering the spin states of dark excitons so that radiative recombination of e-h pairs can be detected. Both conventional and tip-enhanced photoluminescence (TEPL) spectroscopy at room temperature are employed in this study. Interestingly, PDMS nano-patches or Au (or Ag) substrates alone cannot alter the spin states of dark excitons as manifested in our experiments. TEPL mapping with a spatial resolution of < 10 nm on several samples reveals a spatial distribution of dark excitonic emissions confined within PDMS nano-patches when Au or Ag are used as the substrates. Finally, by removing the PDMS nano-patches from the top of $WSe_2$ monolayers we are able to restore the bright excitonic emissions around 1.65 eV.

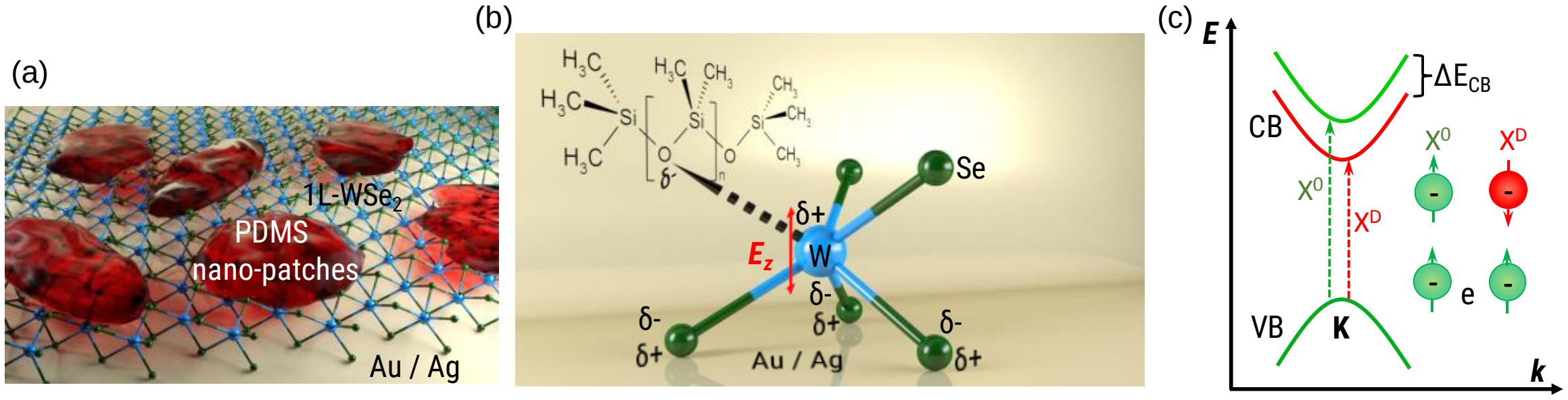

Fig. 1. Scheme of the sample structure, out-of-plane dipole formation and electronic band structure in 1L-$WSe_2$. Schematic of monolayer $WSe_2$ sandwiched by Au (or Ag) and PDMS nano-patches (a). Two-way formation of out-of-plane dipole in $WSe_2$ (b): one via metal substrate (bottom) and other via polar Si-O in PDMS (top). Formation of top dipoles via chalcogen vacancies are the most probable scenario. Spin-assisted optical transitions in monolayer $WSe_2$ (c). Spin up (top) and spin down (bottom) CB bands involved in the bright and dark excitons, respectively.

## Sample preparation

Monolayer $WSe_2$ (exfoliated from bulk $WSe_2$ crystal purchased from HQ Graphene) flakes are transferred on the desired substrate using a conventional deterministic dry transfer method, a commonly adopted technique for 2D monolayers and their heterostack preparation. At first, a monolayer $WSe_2$ is exfoliated on both homemade (preparation details are discussed in the supplementary information, SI-1) and commercially available PDMS (Gel-Film® PF-40-X4 sold by Gel-Pak) film using Nitto Blue tape. After confirming the monolayer thickness of $WSe_2$ using micro PL measurements (see Fig. SI1), the PDMS stamps supported by microscopic glass slides are aligned on top of the desired substrate. In total twenty monolayer samples on Au (or Ag) (thermally evaporated 100 nm film on Si substrate) and four on 300 nm $SiO_2$ substrates are prepared. Among them, fifteen samples are made of homemade PDMS and the remaining nine samples are made using commercial PDMS. All transfer processes are monitored under the optical microscope with a 10x magnification objective. It is generally known that 2D samples prepared by this process contain hydrocarbon including PDMS residuals on the flake surface.[36-37] Therefore, annealing at high temperature is a common practice to clean the 2D surface since these residuals can be mobilised and often segregated into isolated pockets. After the transfer, all samples are annealed at 150°C for 2 hours in an inert atmosphere ($N_2$ chamber, $O_2$ and $H_2O$ < 1 ppm). Fig. 2a displays the atomic force microscopy (AFM) topography images (optical micrographs can be found in Fig. SI2) of one representative sample on Au substrate before and after annealing. As can be seen, before annealing PDMS is hardly visible on the flake since it is distributed rather homogeneously as a thin film on top. However, the phase and the surface potential (SP) images (see SI-3) taken simultaneously on this sample reveal a more heterogeneous surface, which may better indicate the PDMS coverage on the sample. Moreover, comparing the thickness of the flake before and after annealing, it is also evident that the monolayer flake has a higher thickness before the annealing, which most likely originates from residual PDMS. After annealing, a drastic change in the topography is observed, revealing nano-patch formation on the monolayer flake. This is also visible in the corresponding phase and SP images taken simultaneously (SI-3). The dynamics of the formation of such nano-patches after annealing may be related to inherent defects within the $WSe_2$ film. It is well known that TMDCs contain intrinsic defects in the form of chalcogen vacancies.[25] The density of the chalcogen vacancies can further be increased during exfoliation and transfer of the flakes. Since PDMS has a polar chain of Si-O bonds, it is reasonable to assume that during the annealing process PDMS will agglomerate around intrinsic chalcogen vacancies or at grain boundaries. To verify that the nano-patches are formed on top of $WSe_2$ – not agglomerated underneath, we also acquired spatially resolved current maps of our samples using an Au tip in I-Top AFM mode (see methods and SI-4 for details). As can be seen from the current maps (see Fig. SI3), the areas covered by the nano-patches have current values at the noise level (pA) at a bias of 0.05 V; whereas, the clean $WSe_2$ surface show a very good conductivity (4500 pA).

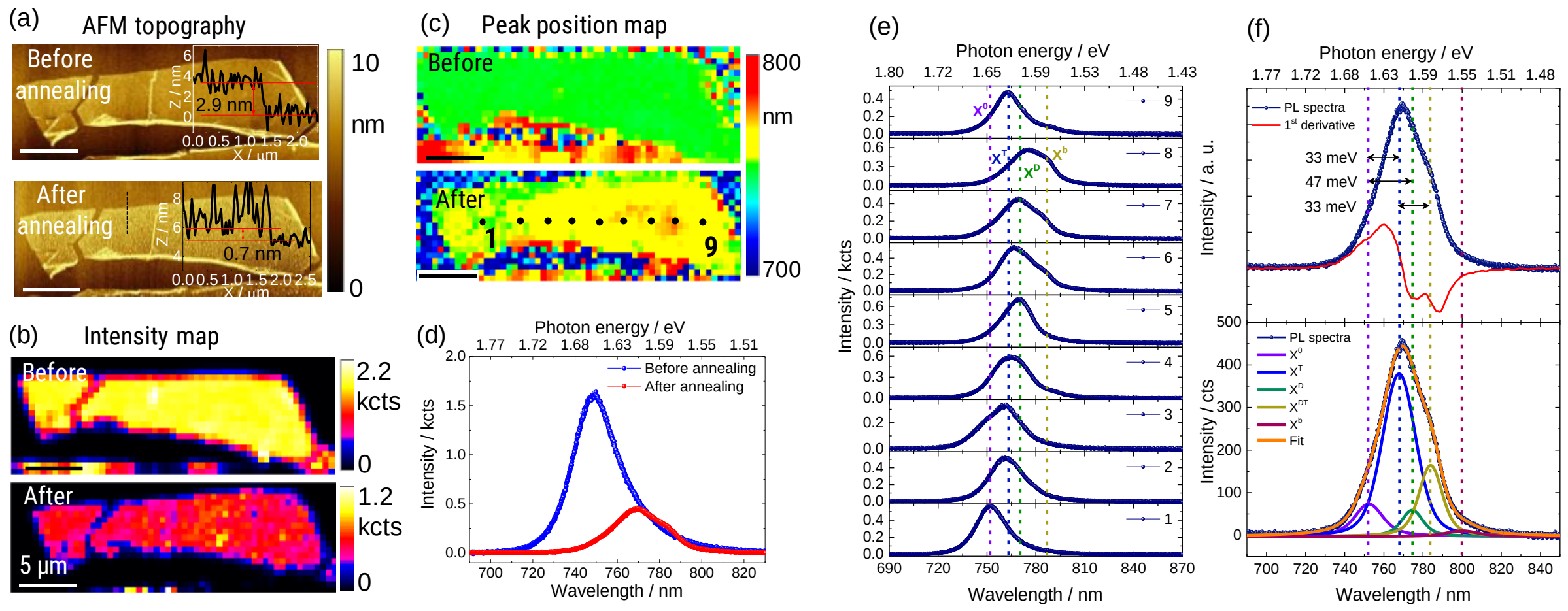

Figure 2. AFM topography and optical characterization of dark excitons at room temperature. AFM topography of 1L-$WSe_2$ before and after annealing deposited on Au (a). The insets are the respective height profiles of the flake taken along the black dotted line shown in bottom image. Nano-patches can be seen after annealing formed by the PDMS residuals. Corresponding optical, phase and SP images are shown in SI-3. Spatial distribution of PL intensity, and peak position map of the flake (b) and (c) before and after annealing, respectively. The PL map was acquired using 100x, 0.9 NA objective and 532 nm excitation within the spectral range of (720 – 800) nm. Both intensity and peak position maps display heterogeneous spatial distributions originating from radiative emission of dark excitons (see main text). The scale bar is 5 µm for all the images. Comparison of PL spectra before and after annealing indicate stronger flake-substrate interaction after annealing (d). A sequence of PL spectra taken on the dots shown in the peak position map (e). The PL spectra clearly indicate the emission from dark excitons. Spectral breakdown of the spectrum 7 shown in (e) together with $1^{st}$ order derivative (f). Apart from neutral excitons and trions we can also observe dark excitons and trions. An additional feature separated by ~50 meV from the dark exciton can also be observed consistently in the spectra. We attributed this peak to the defect bound dark excitonic state.

To confirm the source of these nano-patches conventional and imaging X-ray photoemission spectroscopy (XPS) are performed on one of the samples prepared on Au. The results are presented in SI-5. In the narrow-scan XPS spectra apart from tungsten and selenium of $WSe_2$, carbon, oxygen, and silicon are detected stemming from PDMS. One can see from the Si2s imaging spectra (Fig. SI4a) that the silicon signal in particular coincides with the areas covered by $WSe_2$ flakes. The XPS sampling depth with the Al Ka X-ray source is less than 10 nm. Since the thickness of the Au film is about 100 nm, contributions of Si from the substrate can be completely excluded. Therefore, the only source of Si in our experiments comes from the PDMS residuals.

## Results

**Radiative emission of dark excitons:** Fig. 2b and c display the spatial maps of PL intensity and peak position of the monolayer $WSe_2$ sample shown in Fig. 2a obtained in a micro-PL configuration at room temperature. The PL spectra were acquired using a 100x, 0.9 NA objective and 532 nm continuous wave (CW) laser excitation with a power of 100 μW measured at the sample surface. For the ease of presentation the results obtained for Au substrates are discussed here, while results obtained for the Ag substrates are presented in SI. Before annealing, both the PL intensity and the peak position maps show a homogeneous distribution over the whole flake. Interestingly, after annealing this spatial homogeneity as well as the spectral behaviour alter drastically. As shown in Fig. 2d, before annealing $WSe_2$ has a single feature centred at 750 nm (equivalent to 1.65 eV) characteristic for the emission from the neutral A exciton (bright, $X^0$ emission) in the monolayer. However, once the sample is annealed, the PL intensity decreases dramatically. More importantly, new features appear in the spectra. One of the reasons for the drastic reduction of the PL intensity can be due to a stronger $WSe_2$-Au interaction after annealing.[38] In order to get more spectral information, a series of spectra were taken along the dots shown in Fig. 2c and presented in Fig. 2e. It is evident from Fig. 2e that the spectral weights of the spectra are divergent from each other with at least four distinctive features of different intensity ratios. The first two features can be attributed to the neutral A exciton and trion ($X^0$ and $X^T$ from the upper branch of the CB as shown in Fig. 1c) separated by ~30 meV from each other. The remaining features stem from dark excitons and are the main focus of our discussion in this work. It is important to note that the total of twenty samples prepared on Au and Ag as mentioned above show consistent results. Additionally, we also observed a systematic substrate-to-substrate shift of all the features up to 20 nm (equivalent to 44 meV). We attributed this effect to the local dielectric disorder originating from the substrate morphology and cleanness.[39]

In order to confirm that none of these PL features stems from PDMS nano-patches we also prepared 1L-$MoS_2$ samples following the same procedures (See SI-6). Since 1L-$MoS_2$ does not have any contribution around 780 nm (1.59 eV), any PL band appearing in this range would come from the nano-patches. Therefore, comparing Fig. SI5 before and after annealing we can confirm that the PL features are solely from $WSe_2$ in our samples. We also studied the influence of longer annealing time. For this purpose, two samples were annealed a second time for another two hours in the same inert atmosphere at 150°C. The PL results are presented in SI-7. It can be seen in Fig. SI6 that the radiative emission from dark excitons is still present after the second time annealing, meaning that the annealing time has only little influence in brightening the dark excitons as long as nano-patches are formed on top of the monolayer $WSe_2$.

To get deeper insight in the origin of the features, spectrum 7 in Fig. 2e is deconvoluted using Voigt functions. The deconvoluted spectrum together with 1st order derivative are presented in Fig. 2f. There are four distinctive peaks, which can be identified in the 1st order derivative of the spectra (top part in Fig. 2f). From the fitting (bottom part in Fig. 2f) we can assign these four peaks to the bright exciton $X^0$ at 1.648 eV, the bright trion $X^T$ at 1.615 eV, the dark exciton $X^D$ at 1.601 eV, and the dark trion $X^{DT}$ at 1.582 eV. The energy difference between $X^0$ and $X^D$ is 47 meV and between $X^T$ and $X^{DT}$ 33 meV. This is in very good agreement to the reported values in the literature.[27,28, 32] One notable characteristic of dark excitons is their narrow linewidth compared to the bright excitons.[30, 32, 34] In our case, the typical full width at half maximum (FWHM) of bright excitons is 35-40 meV, whereas the FWHM of dark excitons is (25 ± 5) meV. In addition to these four peaks, we also observe a relatively weak feature consistently for all samples appearing around 50 meV below the dark exciton. The origin of this peak is not clear at the moment. However, we expect that this peak can be assigned to an extrinsic charged defect bound dark excitonic state. The argument behind our assignment is that when a dark exciton is captured by a localized electron or hole, it becomes bright via valley mixing in the conduction band[23]. Moreover, the binding energy of this particular band is too large (~100 meV) to stem from four or five particle complexes[40-43] or from phonon replica of the dark exciton[31] as reported in the literature.

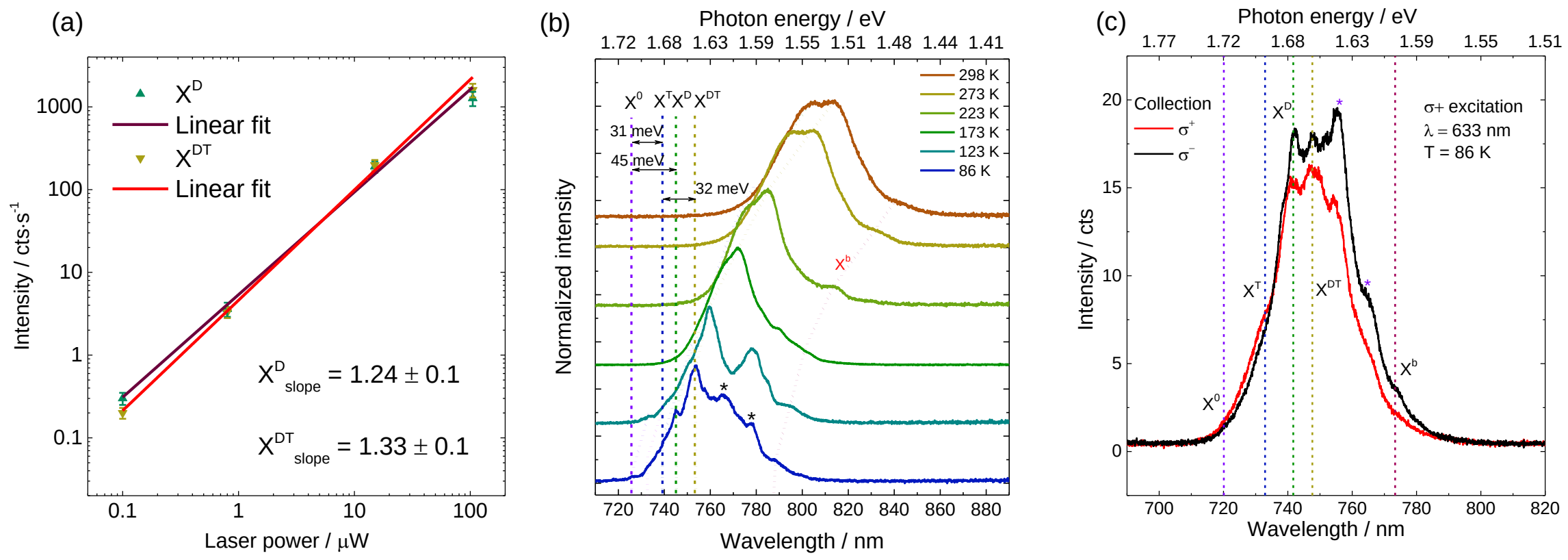

Fig. 3. Probing radiative emission of dark excitons by varying laser power, temperature, and polarisation. Photo-excitation power dependent PL spectra of 1L-$WSe_2$ on Au (a). Both dark exciton and trion exhibit a linear excitation power dependence. Temperature dependent PL of monolayer $WSe_2$ on Au (b). Both (a) and (b) were acquired under 532 nm excitation using 100x, 0.9 NA (a), and 50x, 0.5 NA (b) objectives. A similar study on Ag is shown in the SI-7. Circular polarization dependent PL of monolayer $WSe_2$ (c). $\sigma^-/\sigma^\pm$ (excitation/collection) spectra are shown in SI-8.

To study the photoexcitation power dependent behaviour of dark emissions we recorded PL spectra at different laser power. Since biexcitons in this system have comparable emission energy to dark excitons, our excitation power dependent PL study can clarify such possibility. Fig. 3a presents one typical log plot of photoexcitation power dependent PL intensity of $X^D$ and $X^{DT}$ emissions in our samples. Corresponding PL spectra with fitted peaks are shown in Fig. SI8a. One of the key features of biexcitons is the quadratic response of the emission intensity with excitation power.[44] Hence, from the results shown in Fig. 3a such possibility can be excluded. Interestingly, both $X^D$ and $X^{DT}$ show linear but slightly different power dependence. $X^{DT}$ has a faster power dependence rate of 1.33 cts·s$^{-1}$·µW$^{-1}$; whereas, $X^D$ has a rate of 1.24 cts·s$^{-1}$·µW$^{-1}$ in very good agreement to the literature.[34]

To observe the temperature dependent behaviour of dark excitons and trions, a series of PL spectra at different temperatures were acquired as shown in Fig. 3b. All spectra were measured using a 0.5 NA objective under 532 nm excitation. A similar study on a Ag substrate is presented in Fig. SI8b. At 86 K, the bright exciton $X^0$ and the trion $X^T$ emissions are located at 1.710 eV and 1.679 eV, respectively. In between the bright exciton and the trion band there is another feature located 14 meV below $X^0$, which may originate from biexcitonic emission.[24, 40-43] As the temperature increases, both bright exciton and biexciton disappear within the trion's shoulder. Biexcitonic states eventually thermally depopulate at elevated temperature due to the small binding energy. However, in the case of excitons, we observe strong trion emission consistently throughout the samples. This can be due to a local doping effect via the substrate from the bottom and via the nano-patches from the top. As an example, impurities or charged defects present in the nano-patches can induce unintentional residual doping in $WSe_2$.[45-46] As a consequence, we observe much stronger trions than neutral excitons in our systems. Importantly, apart from the bright exciton and trion, we also observe the dark exciton and trion located at 1.675 eV and 1.647 eV, respectively. The energy differences between $X^0$ and $X^D$ and $X^T$ and $X^{DT}$ also agree well with room temperature PL measurements (see Fig. 3b). Besides, there is one feature located at 1.573 eV at 86 K, which survives even at room temperature. We assign this peak as $X^b$ as discussed above. Additionally, there are a few other emission bands appearing at 86 K (marked by asterisks), which disappear as the temperature increases. The origin of these bands is unclear at the moment. We cannot assign them to localized defect bound states reported in these system since they are observed at cryogenic temperature and thermally dissociate as the temperature increases.[47] In our systems we can observe them up to 173 K, which is beyond the temperature range reported for these bound states. Moreover, the binding energies of these bands are observed to be above 80 meV. Therefore, these bands do also not originate from the Coulomb bound five particles complexes since their binding energy (~ 50 meV)[40-42] is smaller than the features observed in our system (> 80 meV).

We also performed circular polarisation dependent PL to study valley dynamics of dark excitons. Fig. 3c shows the results for right circular polarisation (CP) excitation and right/left CP collection configurations measured using 633 nm photoexcitation at 86 K. The data for left CP excitation are presented in Fig. SI8c. As expected the bright states (exciton and trion) show enhancement with the same CP direction, in agreement with previous reports.[8] Intriguingly, emission from the dark states shows stronger enhancement with opposite CP direction reflecting the distinctive nature of these states. The exact origin behind this behaviour remains under discussion. He *et al.*[48] reported that the hole-rich dark trion ($X^{D^+}$) shows stronger enhancement in cross-polarization geometry; whereas the electron-rich dark trion ($X^{D^-}$) radiates equal intensity at both polarization geometry. However, Zhang *et al.*[28] revealed the opposite CP nature of the dark excitonic species which was also confirmed in another recent report.[30] It is widely speculated that an intervalley bright-dark exciton scattering process gives rise to a higher population of photogenerated electrons in the opposite valley of the lowest CB. Hence we observe a stronger enhancement with opposite CP emission. Additionally, the features marked by asterisks (bands are also highlighted in 3b) are also enhanced with cross hand polarisation. This particular behaviour indicates the involvement of dark characteristics in these bands similar to the situation with $X^b$ as shown in Fig. 3c.

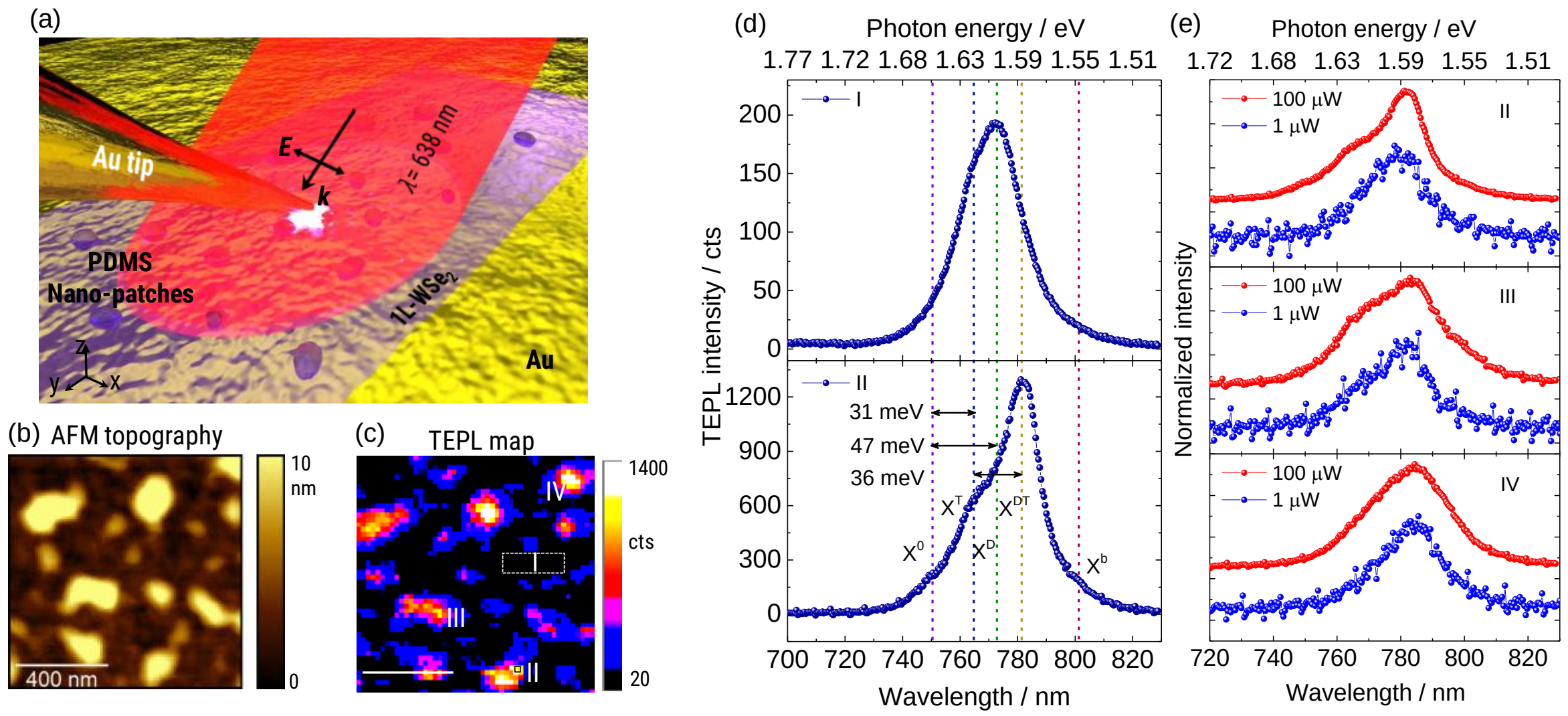

Figure 4. Probing radiative emission of dark excitons in TEPL configuration. Sketch of TEPL experimental configuration (a). AFM topography (b) and corresponding TEPL spatial map (c) of 1L-$WSe_2$ covered by PDMS nano-patches on Au. The TEPL image was created within the spectral range of (740 – 820) nm with a step size 10 nm. Excitation energy was 1.94 eV (638 nm). More TEPL results including those for Ag substrates are presented in SI-9 and SI-10. TEPL spectra of monolayer $WSe_2$ on Au (top), and on PDMS nano-patches (bottom) as highlighted in TEPL map (d). Photoexcitation power dependent TEPL spectra (e).

From the micro-PL measurements, it evident that we observe PL emission from dark excitons and trions at room temperature. Our hypothesis behind this observation is that PDMS nano-patches formed during the annealing process "brighten" the dark excitons. In order to verify this hypothesis, we performed TEPL on our samples with a spatial resolution of <10 nm. Fig. 4 displays one typical TEPL result. More TEPL data can be found in SI-10 and SI-11. The experimental configuration of our TEPL setup is shown schematically in Fig. 4a. All TEPL measurements were performed at a constant tip-sample distance (in this case the tip is in contact with the sample). It is important to note that all TEPL experiments were carried out at room temperature. As can be seen in the AFM topography image (Fig. 4b) nano-patches are randomly distributed on the sample with a typical size from tens to a couple of hundreds of nanometers. The corresponding TEPL map (Fig.4c) has a clear correlation with the AFM image with higher PL intensities coming from the areas covered by nano-patches. The TEPL spectra of a monolayer $WSe_2$ on clean Au (see Fig. 4d top) shows a strong feature around 1.6 eV and two shoulders at the high energy side around 1.63 and 1.65 eV. Comparing these spectra with our micro-PL study (see Fig. 2), the two higher energy features are due to $X^0$ and $X^T$; while the peak around 1.6 eV stems from radiative emission of $X^D$. This is not surprising since we induce an out-of-plane optical dipole moment in our TEPL configuration. Therefore, the orthogonal transition dipole of dark excitons couples to the optical field created at the tip-sample sub-nano gap and enhances the radiative emission of dark excitons. This is in very good agreement with recent $X^D$ observations facilitated by the TEPL geometry at room temperature.[34]

Interestingly, when we optically probe over the nano-patches by an Au tip, the TEPL spectra change dramatically (Fig. 4d bottom) with respect to the spectra recorded on top of Au. First, we can now resolve the dark excitonic emissions much better and secondly, the overall intensity increases by 6 to 10 times. The PL quantum yield (QY) in the low excitation limit for a certain exciton population within the light cone depends on the relative spectral distance between dark and bright states. Being the lowest state of the spin split CB in $WSe_2$, the PL QY in this system therefore depends on the thermal population of the spectrally higher bright states and secondly the radiative emission efficiency of dark excitonic states. This picture can be well understood by the temperature dependent PL QY study in TMDCs since at 0 K all excitons occupy the lowest CB state.[49] In our system, we observed increased QY on top of nano-patches due to both thermal population to the bright states (less efficient due to larger separation ~ 45 meV) and brightening of dark excitons. Hence, we can resolve the $X^D$ related states much better on top of nano-patches. It is important to note that, in our TEPL measurement, we observed $X^D$ and $X^{DT}$ binding energies varying from (45 – 52) meV and (33 – 38) meV respectively. The variation in binding energies from nano-patch to nano-patch can also indicate the variation of local dielectric environment on the same sample originating from dielectric disorder as discussed in a previous report.[39]

Fig. 4e presents excitation power dependent TEPL spectra taken from three different nano-patches as shown in the TEPL map in Fig. 4c. As can be seen the intensity of all spectral features changes in similar proportion excluding any biexciton formation. This is in good agreement with the excitation power dependent micro-PL in Fig. 3a. Interestingly, the spectral weight of each feature in these three spectra are different. This is also true for other nano-patches investigated in this work (see Fig.SI9). The heterogeneity in the spectral weight of these features indicates local doping processes varying from patch to patch.

Since Raman spectroscopy is a powerful technique for investigating local heterogeneities[50] we also recorded tip-enhanced Raman scattering (TERS) spectra. Fig. 5a displays a TEPL map of 1L-$WSe_2$ on Au. The corresponding AFM topography image is shown in SI- 12. Two representative TEPL spectra – one on top of Au (marked by a circle) and one on top of nano-patch (marked by a rectangle) are presented in Fig. 5b. As can be seen, the TEPL spectrum on top of the nano-patch shows more features compared to the spectrum on Au (nano-patch free region). The corresponding TERS spectra of the two regions are displayed in Fig. 5c (top panel). More TERS spectra together with corresponding TEPL are shown in SI-12 and SI-13. Comparing the TERS spectra recorded on a nano-patch to those on Au one can see that the overall spectral weight taken on a nano-patch is shifted to higher wavenumbers. In order to get more information, we deconvoluted both spectra (Fig. 5c middle and bottom panel) using Lorentz functions. Since bulk $WSe_2$ belongs to the $D_{6h}$ point group (monolayer belongs to $D_{3h}$), it has two prominent first order Raman modes with $E_{2g}$ and $A_{1g}$ symmetries (in monolayer becomes $E'$ and $A'$). Even though these two Raman modes are well separated in bulk, owing to opposite layer dependent sensitivity they become degenerate in monolayer as denoted by $E\,(\Gamma) + A\,(\Gamma)$ in Fig. 4f.[51-52] Due to quasi-resonant excitation by the 638 nm laser we can also observe features involving second order phonons below and above the first order Raman modes.[51] The feature around 235 $cm^{-1}$ involves in-plane phonons located inside the interior of the Brillouin zone at the $M$ point ($E\,(M)$). The modes located at 255 $cm^{-1}$ and 260 $cm^{-1}$ are overtones of the LA phonon at $M$ point of the Brillouin zone, *i.e.* $2LA\,(M)$, and a phonon having $A$ symmetry at the $M$ point, *i.e.* $A\,(M)$, respectively. Since $E\,(\Gamma)$ and $A\,(\Gamma)$ are degenerate in monolayer, monitoring these modes for possible local strain or doping is a challenging task. However, the $A\,(M)$ mode is in particular sensitive to disorder (in particular defects) similar to graphene.[51, 53] Therefore, comparing the intensity ratio $IR = \left({}^{A\,(M)}\!/_{A\,(\Gamma)}\right)$ can provide a hint of local relative defect density. In our spectra the intensity ratio on Au, $IR_{Au}$, is determined to be 0.8. The value is homogeneous throughout the scanned area with a standard deviation of 0.09. When measured on a nano-patch, the intensity ratio, $IR_{np}$ is determined to be 1.12, which also varies from patch to patch (see SI-13). The intensity ratio, $IR$ on top of Au and nano-patches is a clear indication that there are more defects of $WSe_2$ under PDMS nano-patches than on Au. This also supports our hypothesis that PDMS nano-patches are formed around defect sites in $WSe_2$ during annealing. Note that *E (M)* is also sensitive to structural disorder. However, due to the weak sensitivity of in-plane modes in the TERS configuration we could not resolve another in-plane mode *E (K)*, which is in the proximity of *E (M)* (see micro-Raman spectra in Fig. SI11e). Hence, the deconvoluted spectral weight of *E (M)* most likely has a contribution from *E (K)* as well. Therefore, we did not monitor this peak for defect concentration in this study.

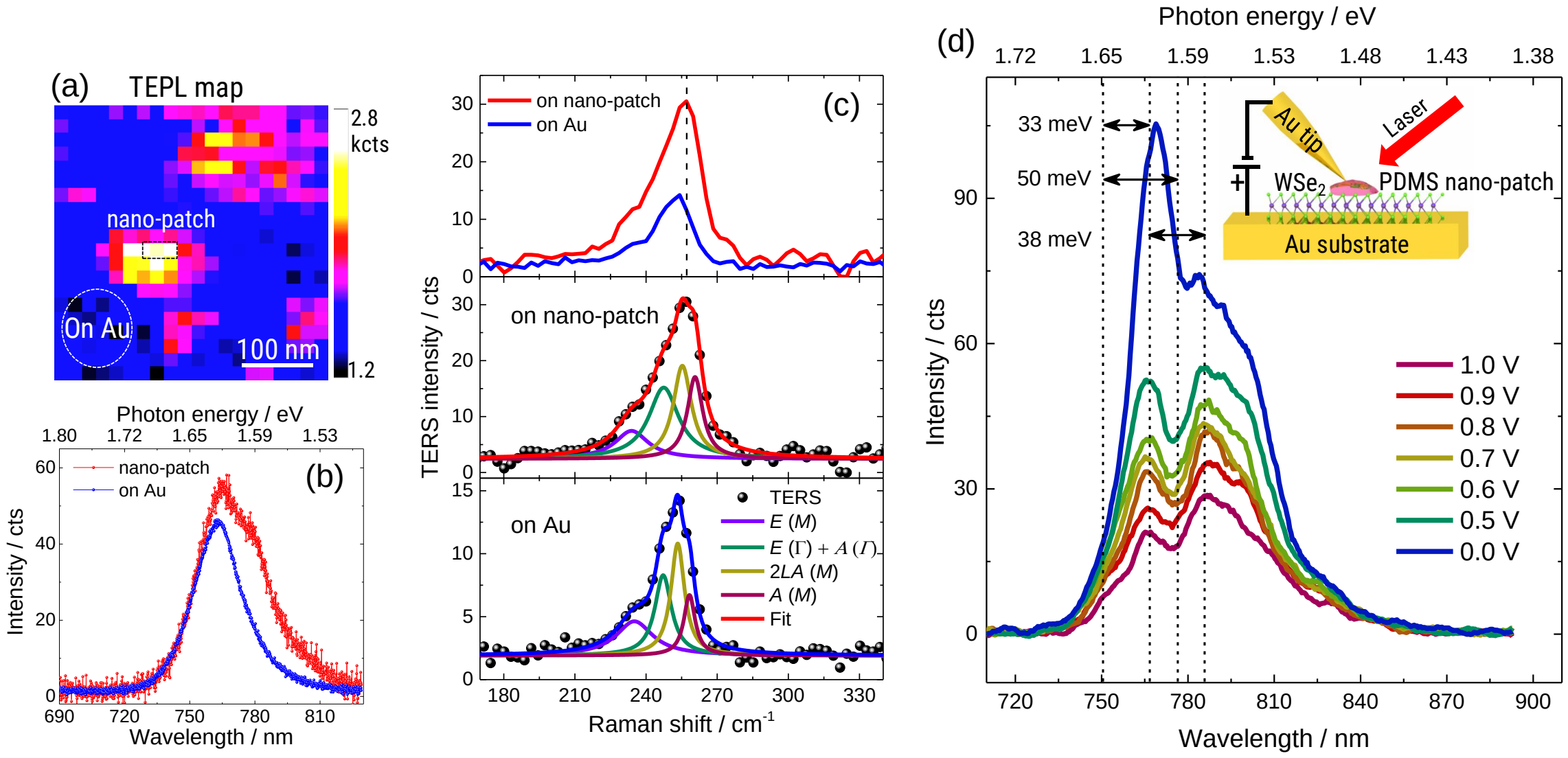

Figure 5. TERS signature of dark excitons and electric field dependent TEPL of 1L-$WSe_2$. TEPL map 1L-$WSe_2$ on Au (a). The AFM topography image of the TEPL map area is presented in SI-13. The TEPL image was created with a step size 10 nm within the spectral range of (740 – 820) nm. Corresponding TEPL (b) and TERS (c) spectra of the two areas marked by a circle (on Au) and a rectangle (on nano-patch) in the TEPL map. More TERS spectra are presented in SI-13 and S-14. The TERS spectra were curve fitted with Voigt functions. Electric field dependent TEPL spectra of 1L-$WSe_2$ on Au (d). The measurement scheme is displayed in the inset. The bias was applied to the sample. TEPL spectra for positive bias are presented here. TEPL spectra for both positive and negative bias are presented in SI-15.

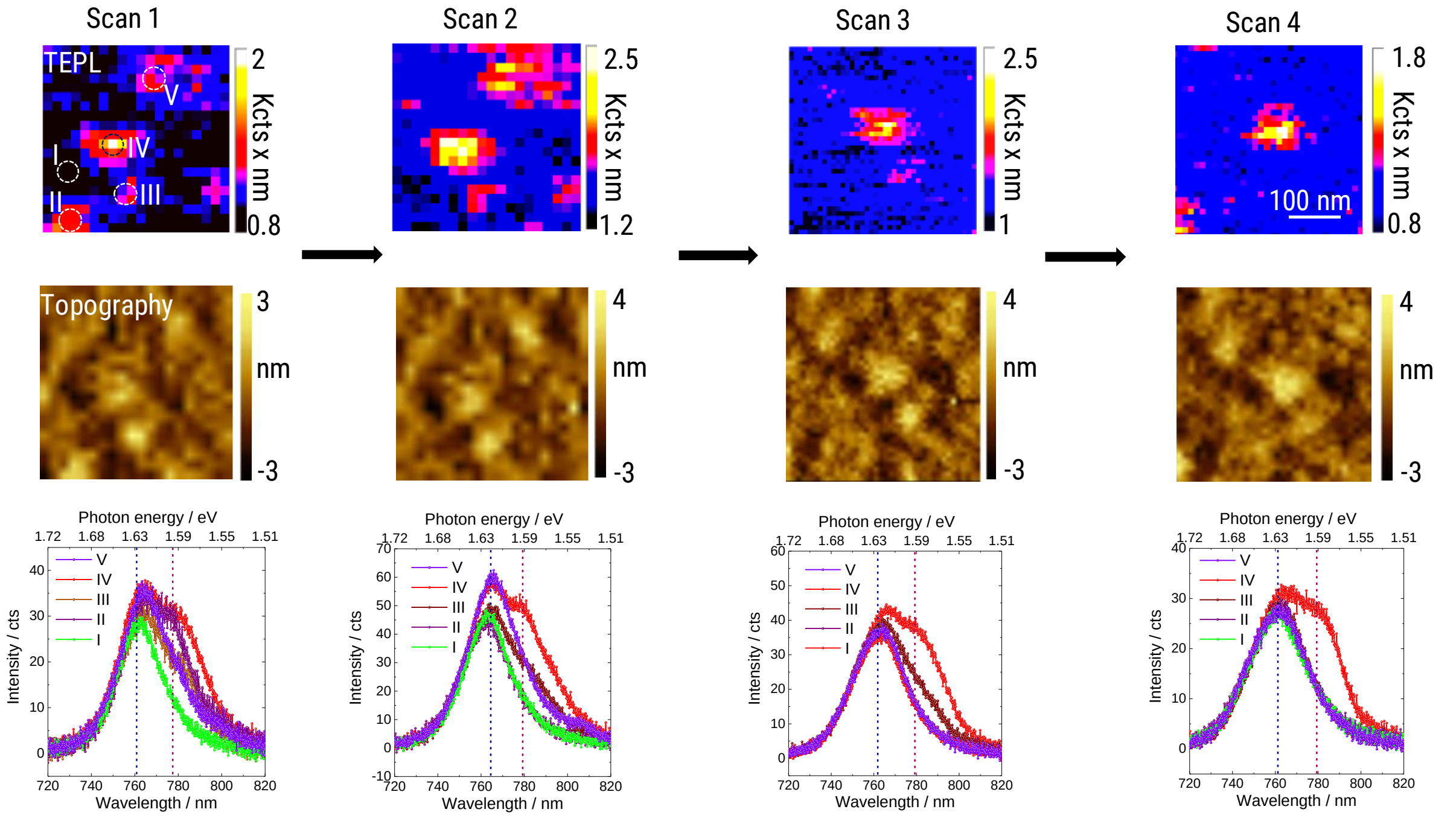

Figure 6. Effect of nano-patches on dark excitonic emission (exciton and trion). Along the columns: TEPL map (top), corresponding AFM topography (middle), and respective TEPL spectra (bottom) of the areas marked by circles in the TEPL map. The first scan started with four nano-patches within the scan area (see topography image). The TEPL spectra on top of all four nano-patches show dark excitonic emissions at the lower energy side of the peak, whereas TEPL on nano-patch-free surface shows weak contribution of $X^D$ only (see text for explanation). In the next three successive scans under illumination ($\lambda$ = 633 nm, 100 µW) three nano-patches are washed off. The corresponding TEPL spectra also change to the nano-patch free TEPL spectra of $WSe_2$.

Since the spin-forbidden dark excitons radiate through an out-of-plane dipole moment, we can also tune these excitonic features via electrostatic gating[29-30, 32]. Fig. 5d presents local gate voltage dependent TEPL of 1L-$WSe_2$ on Au. The experimental configuration of the measurement is shown in the inset of the Fig. 5d. We used a Au tip and the PDMS nano-patches as gate electrode and gate dielectric, respectively. A positive bias was applied to the sample. The illumination/collection geometry of light is the same as stated above for the TEPL measurements. The TEPL spectrum at zero bias shows multiple features including $X^T$, $X^D$, $X^{DT}$, and other features at longer wavelength as discussed above. As the bias increases (hole doping) the intensity of these features decreases gradually – among them $X^T$ (negative bright trion) shows the strongest tunability and the features at the longer wavelength show negligible sensitivity. The intensities of both $X^D$ and $X^{DT}$ also decrease with increasing bias but at a slower rate compared to $X^T$. Interestingly, $X^0$ remains unchanged within the bias voltage range applied. At a sample bias of 1.0 V we can clearly observe all the bright and dark neutral and charged species. Tuning of the PL intensity with a gate voltage is a hallmark of dark excitonic features and agrees well with recent results.[29-30, 32] To confirm that electron doping reverses the situation, we also acquired TEPL spectra at negative sample bias (see Fig. SI13). With the increase of electron concentration (increasing negative bias), the intensity of $X^T$ increases more dramatically than the dark excitonic features confirming the gate tunability of these excitonic species.

As the final part of exploration, we clean the PDMS nano-patches off $WSe_2$ by scanning over it with a Au tip under illumination and performed TEPL to observe $X^D$ emission from the very same spot. Fig. 6 shows a series of AFM scans plus TEPL maps of an area on the sample. After the first scan we can identify four nano-patches within this scanning area as shown in the corresponding AFM topography. Five TEPL spectra – one on Au and four on nano-patches (see TEPL map in Fig. 6) are presented for comparison. After the first scan we can observe stronger dark excitonic emissions on four nano-patches and weaker $X^D$ contribution on Au. After the second scan one of the nano-patches (nano-patch II in Fig.6) is cleaned off. The corresponding TEPL spectra have no evidence of dark excitonic peaks at the lower energy side and match well to the PL on Au (see both TEPL map and spectra in the second column in Fig.6). After the third and fourth scan, nano-patches III and IV are washed off (third and fourth column in Fig. 6). The corresponding TEPL spectra similar to the case of nano-patch II are also not showing dark excitonic shoulders at the lower energy side and become symmetric as the PL on Au. The results clearly indicate that the brightening of dark states originates from the presence of PDMS nano-patches. More importantly, it also demonstrates a simple path to switch dark excitonic emissions in these systems, which will open the door for exploring the rich physics of spin dynamics and possible quantum device applications.

Up to now we demonstrated radiative emission from dark excitonic states in $WSe_2$ monolayers using micro-PL and TEPL at room temperature. From the experimental results it is evident that the brightening of dark excitons is solely confined within the area of PDMS nano-patches created during annealing on the $WSe_2$ monolayer deposited on metallic substrates. Therefore, it is obvious to ask whether a similar behaviour can be observed on other substrates, especially on insulators. To investigate this phenomenon, we prepared four monolayer $WSe_2$ samples on 300 nm $SiO_2$ oxide on Si substrates. Fig. 7a,b display micro-PL intensity and peak position maps of one of the samples prepared on $SiO_2$ substrates before and after annealing. The corresponding PL spectra before and after annealing are shown in Fig. 7c. AFM topography before and

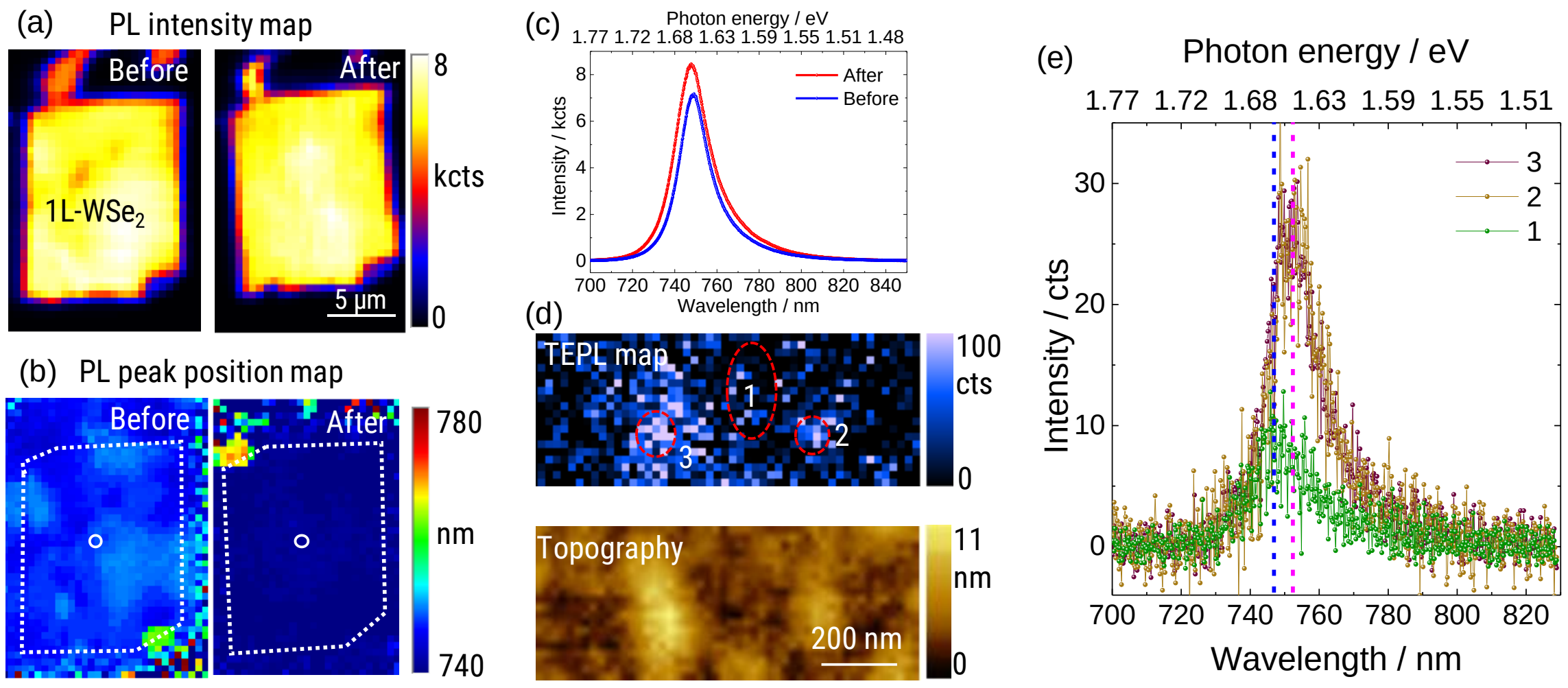

Figue 7. Probing the effect of PDMS nano-patches on excitonic emission in 1L-$WSe_2$ on insulating substrate. Micro-PL intensity (a) and peak position (b) maps of monolayer $WSe_2$ on 300 nm $SiO_2$ substrate before and after annealing. All experimental parameters are the same as micro-PL measurements on metal substrates stated above. White dotted rectangles in the peak position map are a guide for the eyes of monolayer area. Corresponding AFM topography of the flake before and after annealing are presented in SI-15. A scale bar is shown in the PL intensity map. Representative PL spectra before and after annealing taken form the circles shown in the peak position map (c). After annealing no sign of $X^D$ is found in the PL spectra. TEPL map and corresponding AFM topography acquired simultaneously from the sample (d). The scale bar is shown in the topography image. Three spectra are taken from the TEPL map and compared in (e). We observe an enhancement of PL on top of the nano-patches. However, no $X^D$ emission can be seen.

after annealing is presented in SI-15. PL intensity maps before and after annealing show a homogeneous distribution over the whole sample. However, we observe a shift of 7 nm (15 meV) in some areas of the flake in the peak position map before annealing. We attribute this effect to inhomogeneous Coulomb screening due to non-uniform van der Waals coupling between $WSe_2$ and the oxide surface. This is also confirmed by the peak position map after annealing, in which a homogeneous interface is created. Interestingly, both intensity and peak position maps after annealing show homogeneity over the whole flake with the PL peak centred at 750 nm (1.65 eV). This is a sharp contrast to what we observed on Au or Ag substrates. To investigate this in more details, we performed TEPL on this sample as shown in Fig. 7d,e. Like in the case of the metal substrates (see Fig. 4), we observe an increase (though not as efficient) in PL intensity when probed over the nano-patches (Fig. 7d). The corresponding TEPL spectra on top of nano-patches and on $SiO_2$ are presented in Fig. 7e. As can be seen, the TEPL spectra on top of the nano-patches caused, aside from enhancing, also a small red shift of the peak position (~ 6 nm or 13 meV). More importantly, there is hardly any evidence of any dark excitonic emission. The results obtained on $SiO_2$ substrates suggest that the out-of-plane electrostatic field exerted by the PDMS nano-patches on $WSe_2$ is not strong enough to observe brightening of dark excitons at room temperature. The enhanced PL observed on $SiO_2$ is most probably due to more thermal population of the higher bright excitonic states.

**Mechanism of radiative emission of dark exciton:** The above results pointed out the importance of a metallic substrate for the observation of the $X^D$ emission in our samples. Dark exciton radiative emission requires unlocking either spins (intravalley) or large momentum across the valley (intervalley). For the latter case, the assistance from a third body, namely phonon, impurity, or another electron or hole is needed since the exciting photon carries a small amount of momentum. As shown in Fig. 3c, enhanced dark radiative emission with opposite valley polarisation indicates an intervalley scattering process, which does not require a spin flip. However, since we do not observe dark excitons on $SiO_2$ we can safely state that intervalley radiative emission of dark excitons is not the original source in our samples.

Intravalley excitons require spin flip, which can be achieved by an external in-plane magnetic field. Being a member of $D_{3h}$ point group, selection rules dictate that the intensity of an out-of-plane allowed optical transition (in this case radiative dark exciton) is ~ $10^{-3}$ – $10^{-2}$ times the one of the in-plane optical transition (bright exciton) in $WSe_2$.[54] However, a large magnetic field (> 14 T) is required to detect such small radiative dark exciton emission by the Zeeman effect.[27-28] On the contrary, the out-of-plane selection rule is broken by disturbing the reflection symmetry in the surface normal direction. This can be achieved by changing the dielectric environment of the two sides of the monolayer or applying an electric field perpendicular to the plane. This phenomenon facilitates spin flip by a virtual transition in the CB attributed to the SOC mixing and inducing an out-of-plane transition dipole perturbed by the applied field (also known as Bychkov-Rashba effect).[55] In the case of PDMS nano-patches the methoxy group terminated Si-O bond can induce a point charge based out-of-plane dipole moment of up to 1.2 Debye at a distance of 5 Å.[56-57] The resulting electric field value is comparable to the crude estimate of 0.1 V/Å by Slobodeniuk *et al.* on an insulator with vacuum above.[54] Still, at this electric field the radiative decay rate of the dark exciton is negligible compared to that of bright excitons. This probably explains why we do not observe dark-exciton related emission on $SiO_2$ substrate at room temperature. However, using a metal substrate changes the scenario since Au or Ag can form an induced electrostatic dipole at the interface. In addition, both Au and Ag can take surplus charges from $WSe_2$ created e.g. by Se vacancies. This can lead to a stronger interaction between the Si-O bond in the nano-patch and W via Se vacancies as shown in Fig. 1b. More importantly, as a consequence the distance between the Si-O bond to W may decrease. Since the point dipole induced by the Si-O bond exponentially increases with decreasing distance to the probing molecule, an enhanced electrostatic field can be exerted on $WSe_2$.[57] Thus, it can act as a local electrostatic gate and induce an increased population of dark excitons at room temperature. Additionally, as shown in Fig. 4f and SI-12 and SI-13 nano-patches are centred at vacancies, which may further facilitate the breaking of the out-of-plane selection rules in $WSe_2$. More importantly, there is a direct correlation between defects density and radiative emission of dark excitons in our samples. This further strengthens our hypothesis that, due to Se vacancies, we can have a large local electrostatic gating effect via the combination of polar Si-O bond in the nano-patches and the metal substrate. However, more experiments and theoretical studies are required to fully understand the physics behind this behaviour. Finally, similar to the work by Zhou *et al*., a surface plasmon polariton on the metal can also induce a local field to enhance the radiative decay of dark excitons in our system.[32]

## Conclusion

In summary, we demonstrated the radiative emission of dark exciton and trion states in $WSe_2$ monolayer at room temperature using both micro, and nano-PL measurements. Our approach provides a simple way of detecting and manipulating dark excitons over a wide range of temperatures. Apart from $X^D$ and $X^{DT}$, we also observe an additional feature around 50 meV below the dark exciton. This particular feature has never been reported in the literature so far. We assign this feature to an extrinsic impurity state bound to dark exciton state. Micro-PL measurements show that the dark excitonic emissions are homogeneously distributed over the whole flake. While nano-PL measurements with spatial resolution < 10 nm reveal dark excitonic emissions are confined within the PDMS nano-patches on Au or Ag substrates. We attributed this phenomenon to local electrostatic gating via the Si-O bond in the PDMS nano-patch aided by metal substrate. Moreover, we also observe a correlation between the defect concentration and $X^D$ emissions in our samples. We could also tune these excitonic features via local electrostatic gating, which is a hallmark of dark excitons. Finally, by removing the nano-patches from top of $WSe_2$ we were able to tune back the bright excitons. We believe that our results will stimulate more experimental and theoretical work to explore the rich spin physics in the valley of dark exciton landscape. This work is a significant step forward in understanding the physics of dark excitons, which will open the door for the potential application in nano-optics, Qbits, or spintronics.

## Methods

**Micro-PL measurements:** All the micro-PL measurements were performed using a Horiba Xplora Plus equipped with a spectrometer containing 600 l/mm grating and an electron multiplying CCD (EMCCD). A DPSS 532 nm CW laser source was used to excite the samples with an excitation power of 100 µW measured at the sample surface focused by a 100x, 0.9 NA objective. All PL spatial maps were acquired with a step size of 500 x 500 nm$^2$. All PL spectra were corrected with respect to the EMCCD response curve for accurate identification of all PL spectral band features.

Photoexcitation power dependent PL measurements were carried out using the same spectrometer with varying the laser power from 100 nW – 100 µW. For each excitation power a full or line spatial maps on the flake were acquired to reveal the complete picture of the excitation power dependence.

Temperature dependent PL measurements were performed in a liquid nitrogen cooled Linkam stage with a temperature accuracy of 0.1°C using the same spectrometer. A 50x, 0.5 NA long working distance objective was used to excite and collect the PL signal.

Circular polarisation dependent PL emissions were studied using Horiba LabRam HR800 spectrometer containing 600 l/mm grating and an EMCCD detector. A He-Ne 632.8 nm CW laser was focused onto the sample using a 50x, 0.5 NA objective. The temperature of the measurements was set to 86 K using the same Linkam stage mentioned above.

**Nano-PL measurements:** TEPL measurements were performed using the Horiba NanoRaman platform consisting of an AFM (SmartSPM) and Xplora Plus Spectrometer in side illumination geometry. Excitation and collection of photons were realized in the same optical path using a 100x, 0.7 NA long working distance objective at an angle of 65° from the normal to the 2D plane. A p-polarized 638 nm solid state laser was used to excite the sample with a laser power of ~100 µW. An Au tip purchased from Horiba Scientific was used in the experiments. In between the measurements the tip was oscillating in intermittent contact (IC) mode while during the measurements the tip was in contact with the sample for a holding time equal to the TEPL accusation time of 0.2 s.

**TERS measurements:** Experimental configurations of TERS are the same as TEPL measurements. The 600 l/mm grating was used to disperse the signal onto EMCCD. The spectral resolution of the system in this configuration is 6 cm$^{-1}$.

**Electric field dependent TEPL:** For electric field dependent TEPL measurements, PDMS nano-patches were used as the gate dielectric and the Au tip was used as the top electrode. TEPL measurement geometry is the same as stated above. The bias was applied to the sample. The measurements started from the zero bias and was then gradually increased to the positive direction. After that, the negative biases were applied gradually. At the end a zero bias spectrum was acquired in the same run for comparison for all the measurements. In total 10 measurement runs were acquired and the same procedure was followed for all runs.

**SPM measurements:** All SPM measurements were performed using an AIST-NT SPM setup. AFM images were taken using commercially available Si cantilevers in IC mode. KPFM images were acquired using commercially available Pt/Ir tips calibrated against freshly cleaved HOPG substrate. For current images Au coated commercial Si tips of high force constant (Au coating was performed by thermal evaporation using home facility) were used. As shown, one can modify the PDMS nano-patches by repeated AFM scanning over the sample in contact mode. Therefore, in order to acquire current images of the sample we operated the system in IC mode while controlling the tip-sample distance during the local current measurements. In this process, the tip was oscillating at IC frequency while hopping from point to point so that it did not modify or destroy the nano-patches. However, at every measuring point the tip was forced to contact the sample by controlling the set point or nominal force acting on the cantilever. After a few trial and error measurements the optimal force was set on the tip, which confirmed sufficient contact between the tip and the sample for current mapping and at the same time not modifying the nano-patches.

**X-ray photoemission (XPS) measurements:** XPS was performed using a Thermo Scientific ESCALAB 250Xi spectrometer equipped with a monochromatized Al Kα X-ray source (hν = 1486.68 eV). Narrow-scan XPS spectra were acquired form a spot-size of 500 x 500 $\mu m^2$ with an array channeltron detector at a pass energy of 40 eV. XPS imaging was performed in a parallel electron optics mode with a 2D detector at a pass energy of 150 eV.

## Author’s information

All the correspondence and material requests should be addressed to corresponding author:

*Mahfujur Rahaman

Email: mahfujur.rahaman@physik.tu-chemnitz.de

## Acknowledgement

The authors gratefully acknowledge the financial support from the DFG for project ZA 146/44-1, for compute resources via grant INST 270/290-1 FUGB and via the Helmholtz-Kolleg NanoNet.

# Supplementary Information

## Radiative Decay of Dark Exciton Related Emission in a Sandwiched Monolayer $WSe_2$ Revealed by Room Temperature Micro and Nano Photoluminescence

Mahfujur Rahaman[1]*, Oleksandr Selyshchev[1], Yang Pan[1], Ilya Milekhin[1], Apoorva Sharma[1], Georgeta Salvan[1,2], Sibylle Gemming[1,2], Dietrich R T Zahn[1,2]

[1] Institute of Physics, Chemnitz University of Technology, 09107 Chemnitz, Germany
[2] Center for Materials, Architectures and Integration of Nanomembranes (MAIN), Chemnitz University of Technology, 09107 Chemnitz, Germany

Email address: mahfujur.rahaman@physik.tu-chemnitz.de

### SI-1: PDMS preparation

PDMS (Sylgard 184, Dow Corning) elastomer and curing agent was mixed in 10:1 ratio in a clean Petri dish (Petri dish was clean in an ultrasonic bath with acetone x ethanol x DI water for 10 x 15 x 20 mins). After stirring the mixture properly it was placed in a vacuum desiccator for three hours for degassing. After that, the mixture was annealed at 70°C for one hour.

### SI-2: Monolayer $WSe_2$ identification

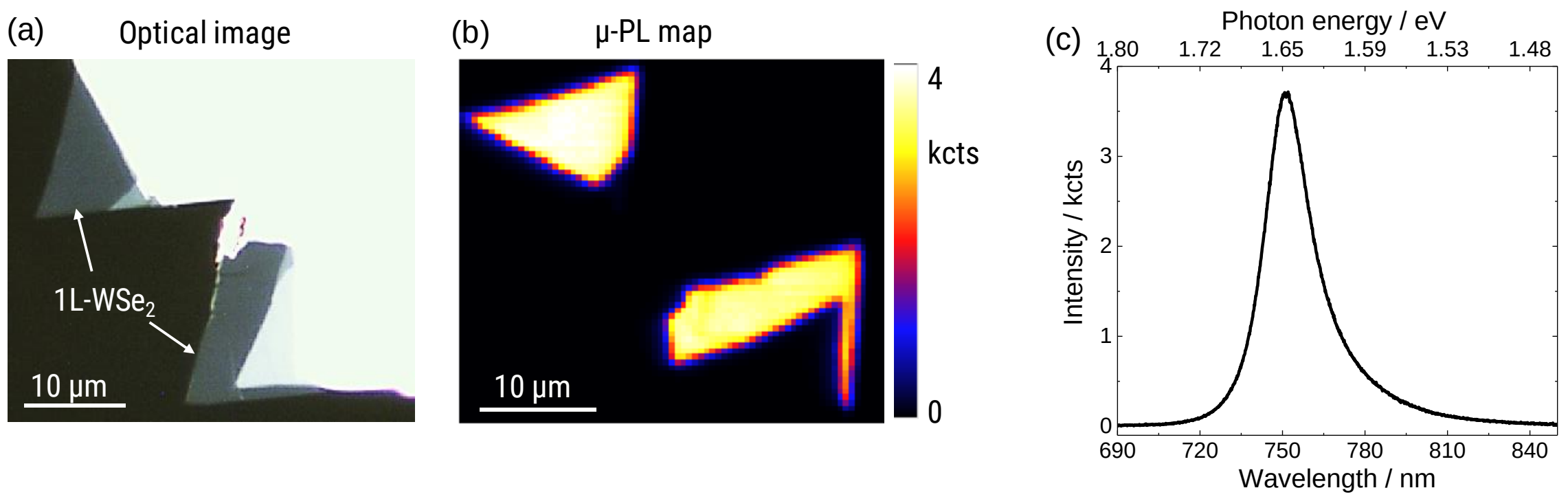

Figure SI1. Optical image of monolayer $WSe_2$ stamped on PDMS (a). Micro-PL map of the monolayer flake on PDMS (b). One representative PL spectrum of monolayer $WSe_2$ taken from the PL map showing PL peak of monolayer $WSe_2$ at 1.65 eV(c).

## SI-3: SPM images of monolayer $WSe_2$ shown in Fig. 2

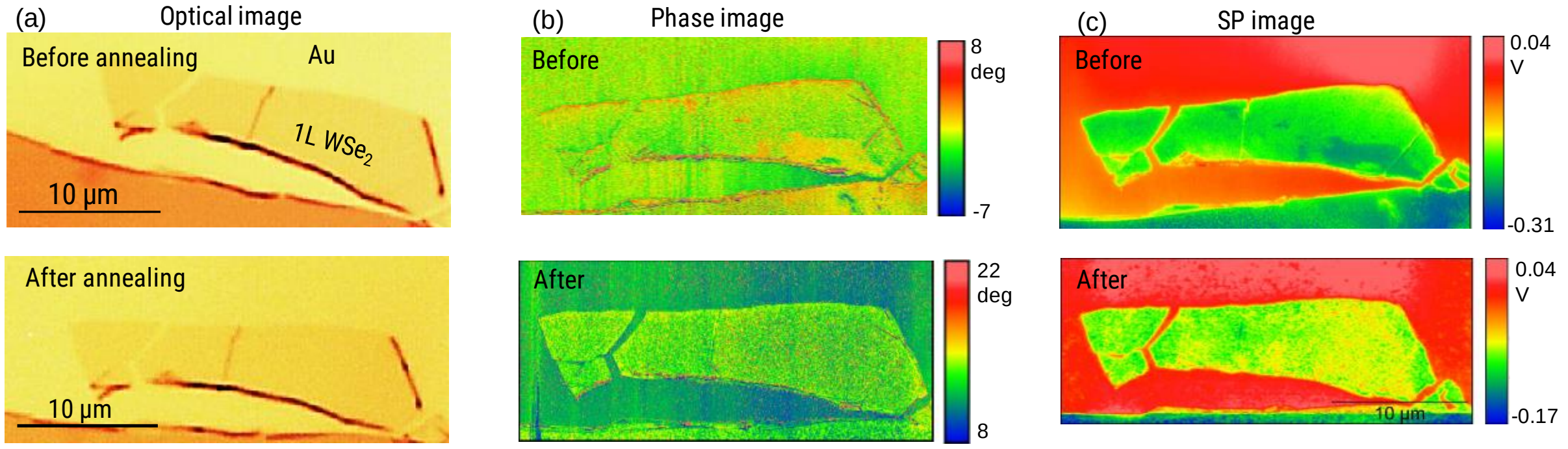

Figure SI2. Optical images of monolayer $WSe_2$ on Au shown in Fig. 2 in main text (a). No change is visible in the optical images before and after annealing. However, both phase (b) and SP (c) images show drastic changes representative of local environment before and after annealing. An AIST-NT SmartSPM set up was used to for the measurements.

## SI-4: Local current imaging on monolayer $WSe_2$ on Au

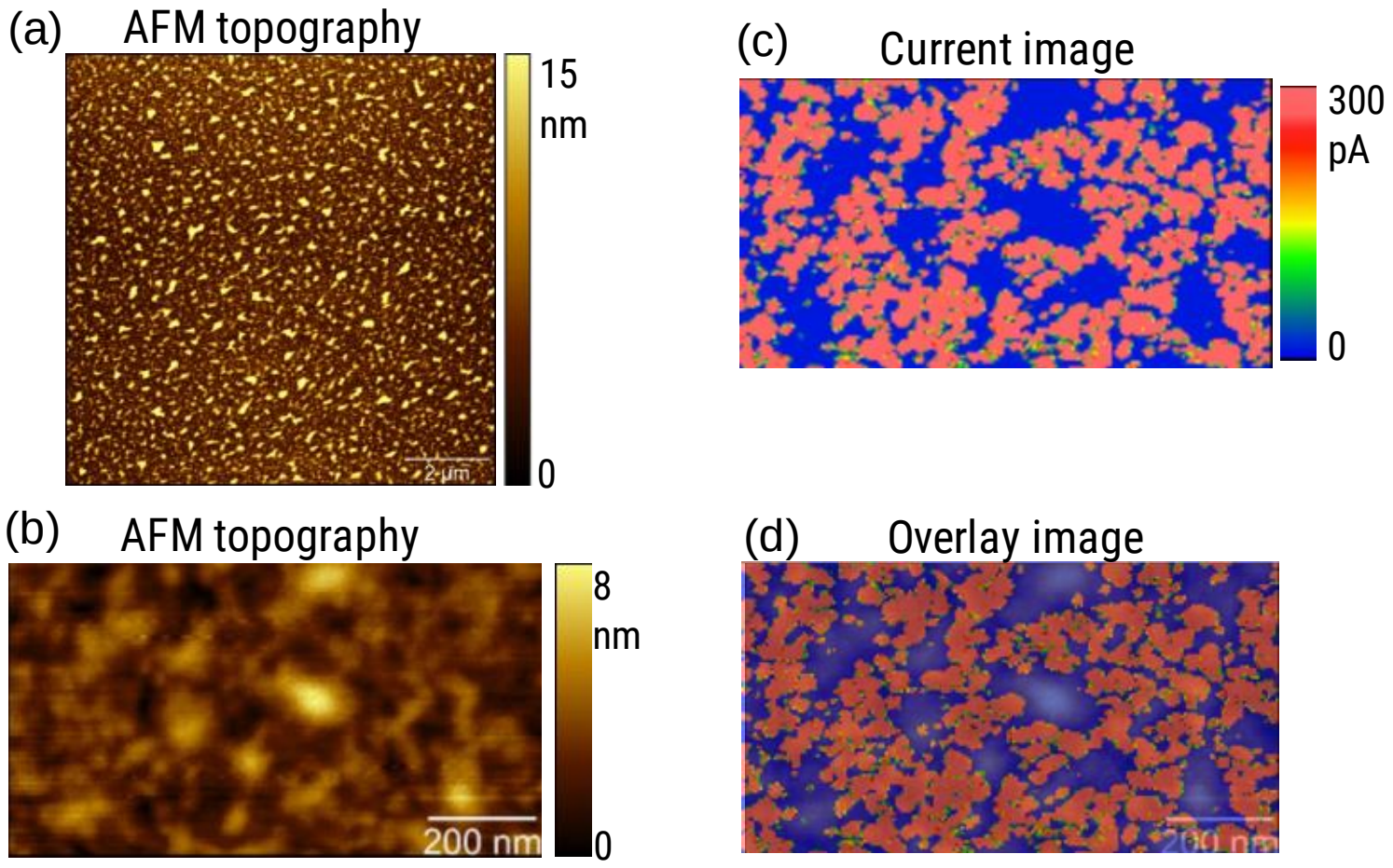

Figure SI3. A typical AFM topography image of 1L-$WSe_2$ on metal substrates after annealing showing nano-patches formation (a). In order to confirm that the nano-patches are formed on top of $WSe_2$ not underneath (at the metal-flake interface) we performed local current imaging on this sample. AFM topography (b) and respective current image (c) and overlay image (d) of one scanning area on the sample. To protect the tip and the nano-patches we operated the AIST-NT SPM system in intermittent contact (IC) mode. During each measuring point the tip was in contact with the sample, while it was hopping between scanning points at near resonant frequency. The tip-sample contact distance was controlled by the force exerted on the cantilever. The force was gradually increased to determine the optimal force, at which we can safely perform local current mapping without damaging the tip or the nano-patches. The final operating parameters are bias voltage = 0.05 V, scanning speed = 0.6 line/s, and step size = 5 nm. As can be seen from (b) to (c), we observe a high current while probing $WSe_2$ on Au and no current when the tip was on top of nano-patches. Since the nano-patches have a height (5-8) nm (higher than $WSe_2$ on Au) any contact related issue can be excluded. Importantly, if the nano-patches were formed at the Au-$WSe_2$ interface then we would still see comparable current over the sample since a conducting path for charges would always be present. However, observing no current indicates the insulating nature of the PDMS nano-patches, which are thus located on top of the flake.

## SI-5: XPS imaging of $WSe_2$ on Au

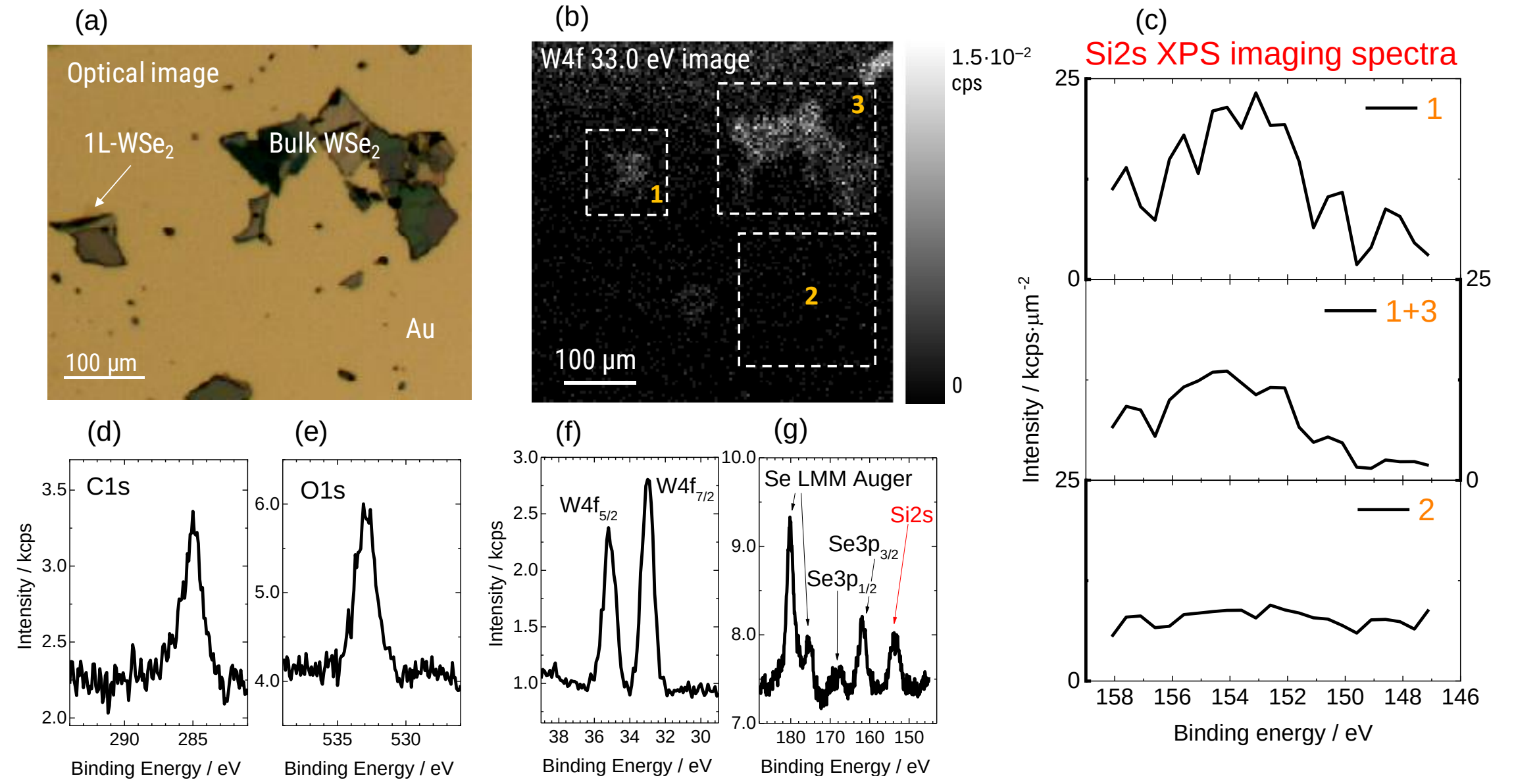

Figure SI4a. Optical image of the $WSe_2$ sample used for the XPS measurements (a). XPS image of the sample area shown in (a) acquired at 33.0 eV representative of $W4f_{7/2}$ core level (b). From the XPS image we can distinguish the $WSe_2$ flakes shown in (a). Spectral imaging was performed within this map area in the spectra range (147-158) eV representative of Si2s core level. Three corresponding spectra averaged over the areas shown in (b) are presented in (c). All three spectra were normalized with respect to the area, over which it was collected. Middle spectrum is the averaged over area 1 and 3. Conventional C1s (d), O1s (e), W4f (f), and Se Auger, Se3p, and Si2s (g) XPS spectra measured within the scanned area shown in (a) and (b) with the 500 x 500 $\mu m^2$ X-ray spot. On $WSe_2$ flake free surface we do not observe any Si contribution. This is expected since the thickness of Au film is 100 nm and the penetration depth of the XPS information is about 10 nm. However, on top of $WSe_2$ flakes we observed a feature around 154 eV corresponding to the Si2s core level. As discussed in the main text, PDMS nano-patches contain methoxy terminated polar Si-O bonds. Therefore, we can conclude that the Si contribution from the flakes are coming from the PDMS nano-patches.

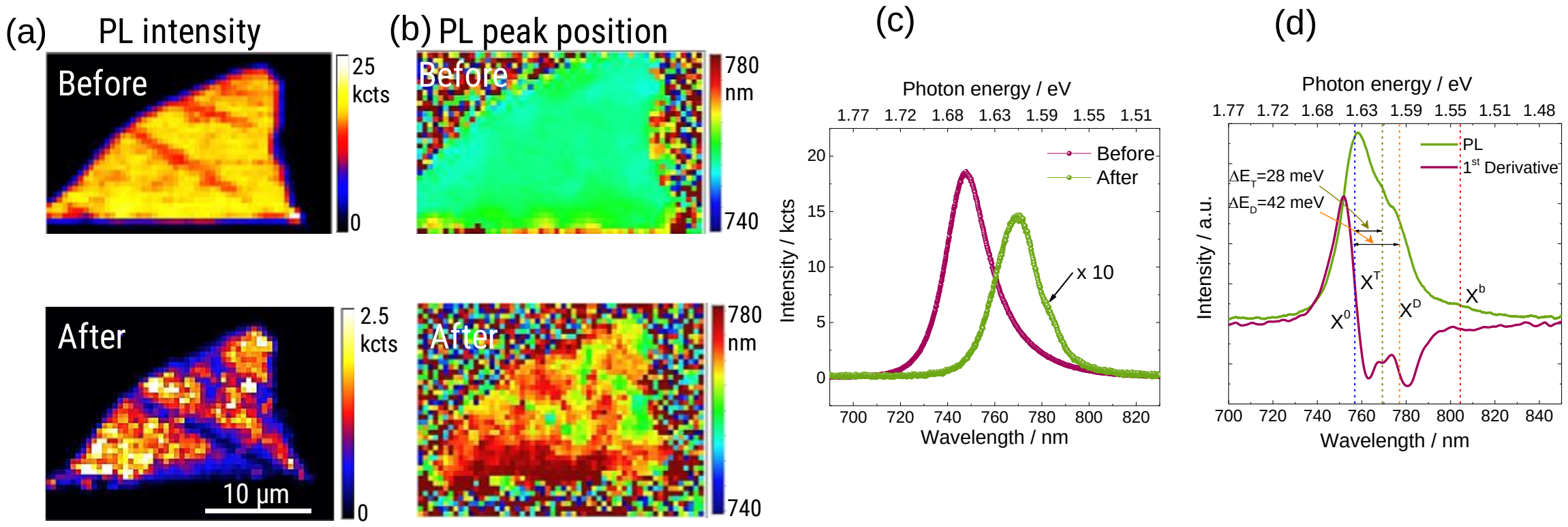

Figure SI4b. PL intensity (a) and peak position map (b) of 1L-$WSe_2$ on Au shown in the optical image in SI4a before and after annealing. PL measurements were performed at room temperature. The PL spectra dramatically changed due to nano-patches formation after annealing. Comparison of the PL spectra before and after annealing (c) and the same PL spectra after annealing shown in (c) with 1$^{st}$ order derivative (d). We can clearly identify the $X^D$ emission in the spectra at room temperature separated by 42 meV from $X^0$.

## SI-6: PL study of 1L-$MoS_2$ on Au

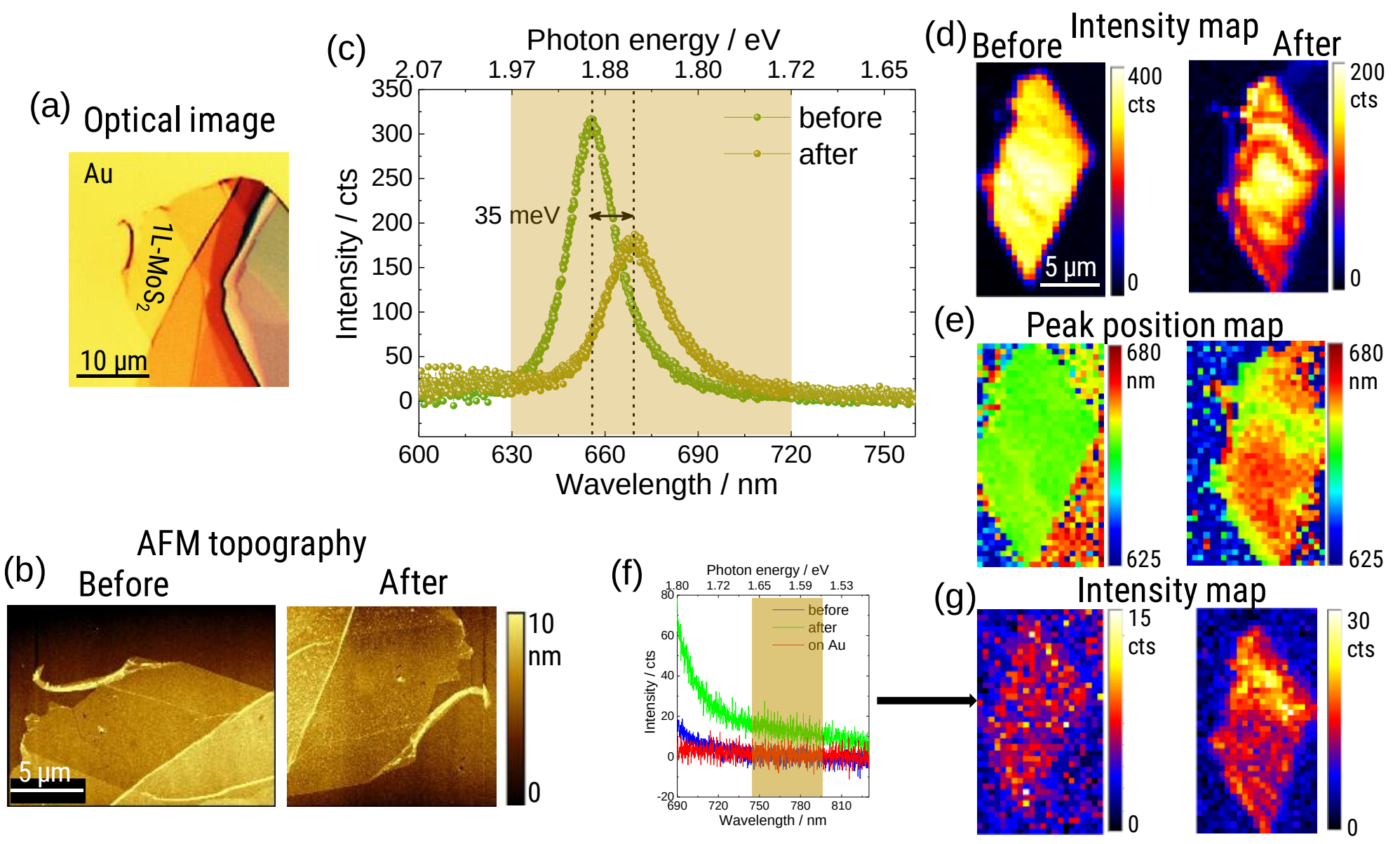

Figure SI5. Optical image of the monolayer $MoS_2$ on Au (a). AFM topography before and after annealing showing nano-patch formation during the annealing process (b). Comparison of PL spectra before and after annealing (c). After annealing we observe a decrease of the PL intensity and peak position shifting to longer wavelength. The decrease of PL intensity suggests a stronger Au-$MoS_2$ interaction. The PL peak position is shifted to lower energy by a value of 35 meV equivalent to the trion binding energy in $MoS_2$. Thus the peak position shifting indicates a stronger contribution from trion and weaker contribution from neutral exciton. PL intensity (d) and peak position (e) map of $MoS_2$ created within the spectral range highlighted in (c). Interestingly, after annealing the PL peak shifted by 35 meV as mentioned and stayed uniform throughout the flake suggesting homogeneous doping induced by the metal substrate. This is expected since a uniform clean interface can be created by annealing. To verify if the PDMS nano-patchescause the PL features around 780 nm, we also acquired map in this spectral region for this sample. PL spectra (f) and spatial map (g) within the highlighted region showing an increase of background originated from the long tail of PL emission in $MoS_2$ after annealing. Since we do not observe any PL feature around 780 nm we can conclude that PL features around 780 nm observed in $WSe_2$ samples originate from $WSe_2$.

## SI-7: Effect of longer annealing on radiative emission of $X^D$

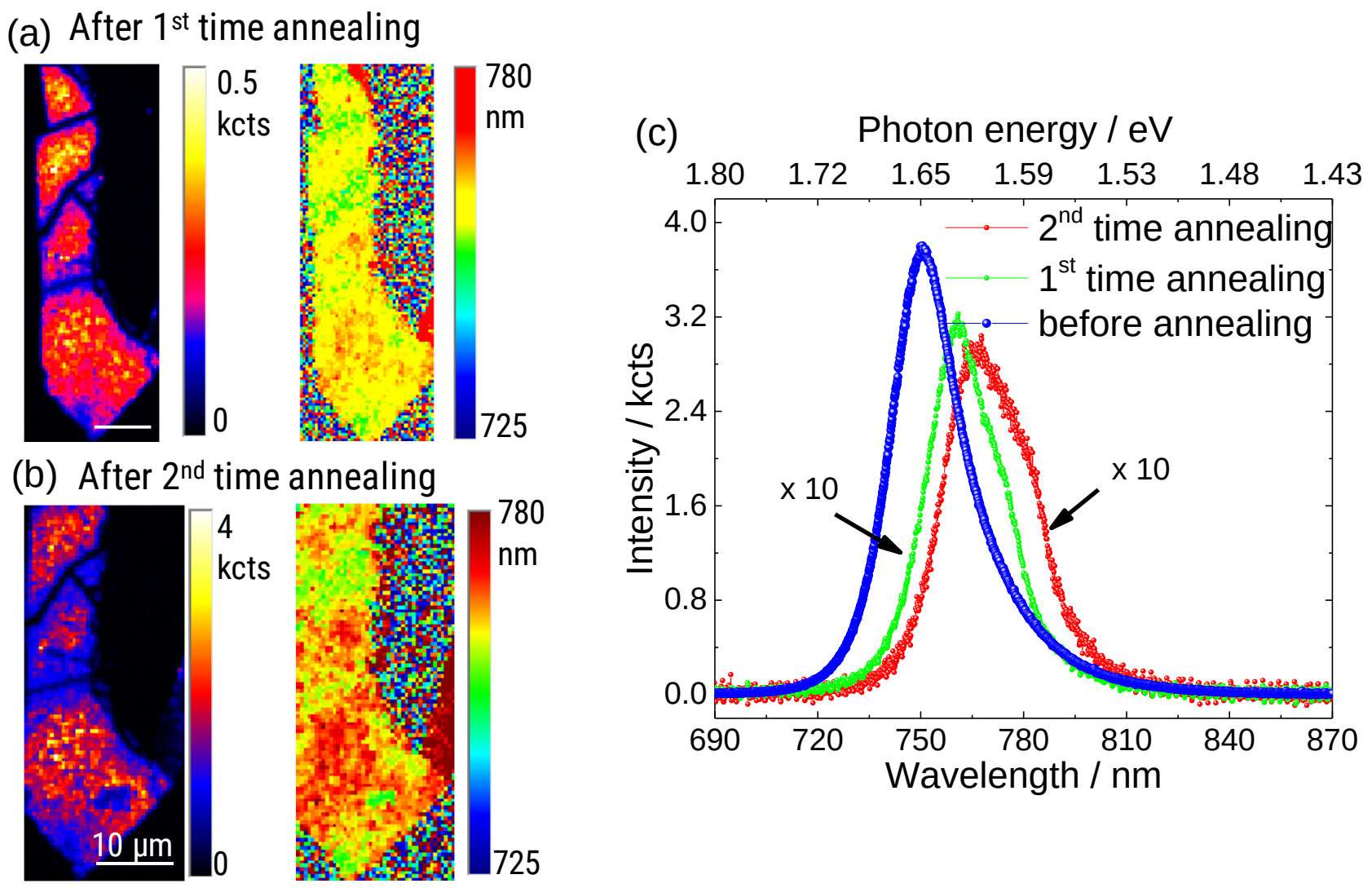

Figure SI6. PL intensity and peak position maps of one 1L-$WSe_2$ on Au after first time annealing (a) and after second time annealing (b). As can be seen both maps show non-uniform distribution of PL peaks of $WSe_2$ originating from radiative emission of $X^D$. The comparison of PL of the same sample before annealing, after 1st time annealing, and after 2nd time annealing is presented in (c). After the first annealing process, the PL intensity decreased drastically. However, after the second annealing the decrease in intensity is negligible suggesting an efficient interface already been created after first annealing. Importantly, in both cases, we observed dark excitonic emissions – though after the second time annealing, the spectral weights of these emissions became more pronounced indicating stronger interaction between PDMS nano-patches and 1L-$WSe_2$. However, the presence of dark excitons in both cases, indicated that annealing has little influence in terms of radiative emission of dark excitons as long as the nano-patches are formed and interact with $WSe_2$.

## S-8: Radiative emission of $X^D$ on Ag substrate

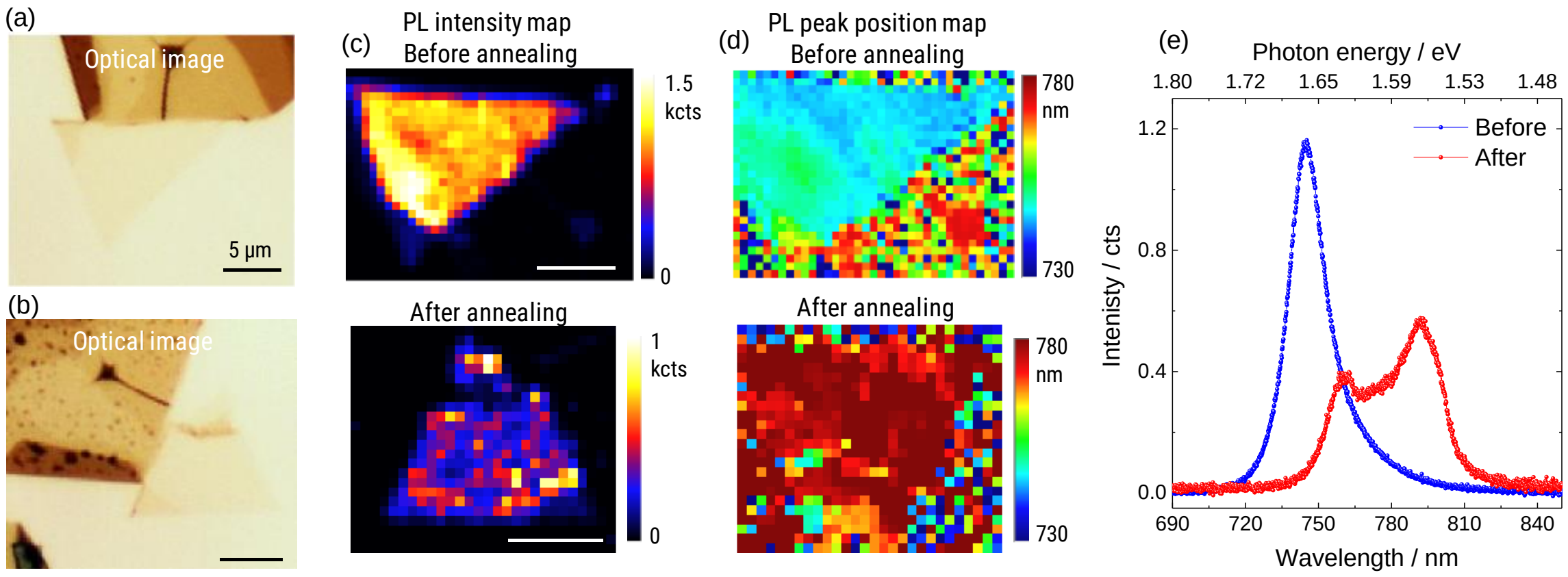

Figure SI7. Optical images of 1L-$WSe_2$ on Ag before and after annealing (a) and (b). PL intensity map (c) and peak position map (d) of the same flake before and after annealing. For comparison representative PL spectra before and after annealing are shown in (e). From the PL results, it is evident that we observe dark excitonic emissions from $WSe_2$ on Ag as well.

## S-9: Photoexcitation power, temperature, and polarization dependent study of dark excitons

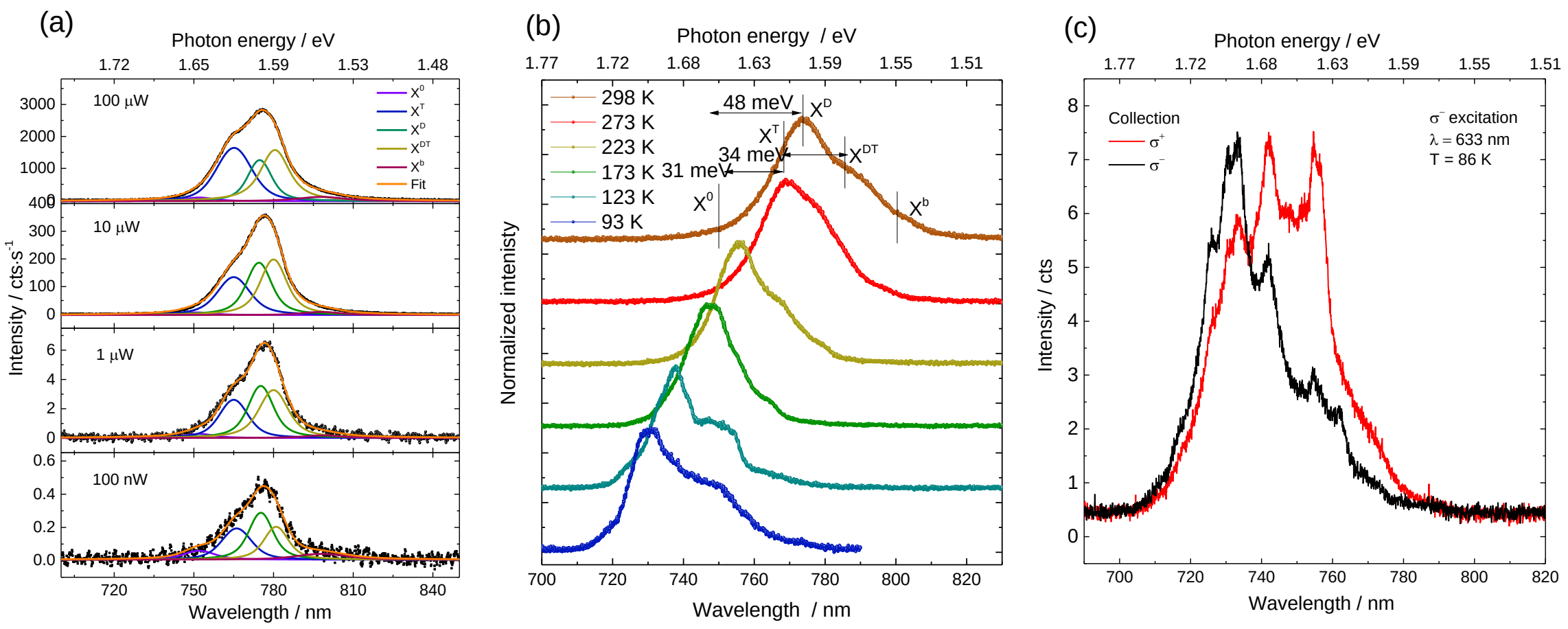

Figure SI8. Photoexcitation power dependent PL of 1L-$WSe_2$ on Au. The measurements were performed using Horiba Xplora Plus spectrometer under 532 nm laser illumination focused with a 100x, 0.9 NA objective. Similar to Fig. 3a, both $X^0$ and $X^T$ have linear power dependence. Temperature dependent PL of 1L-$WSe_2$ on Ag. The measurements were performed using Horiba Labram under 633 nm excitation and an objective of 50x, 0.5 NA. The laser power at the objective surface was 167 µW. A liquid nitrogen cooled Linkam stage was used to control the temperature. The energy difference between $X^0$ and $X^T$, $X^0$ and $X^D$, and $X^T$ and $X^{DT}$ are 31 meV, 48 meV, and 34 meV, respectively, consistent in our work. Left circular polarization dependent PL of $WSe_2$ sample discussed in the main text. With opposite handed polarization we observed enhanced PL from the dark emissions in agreement to the right handed polarization shown in Fig. 3c.

## S-10: TEPL study of 1L-$WSe_2$ on Au

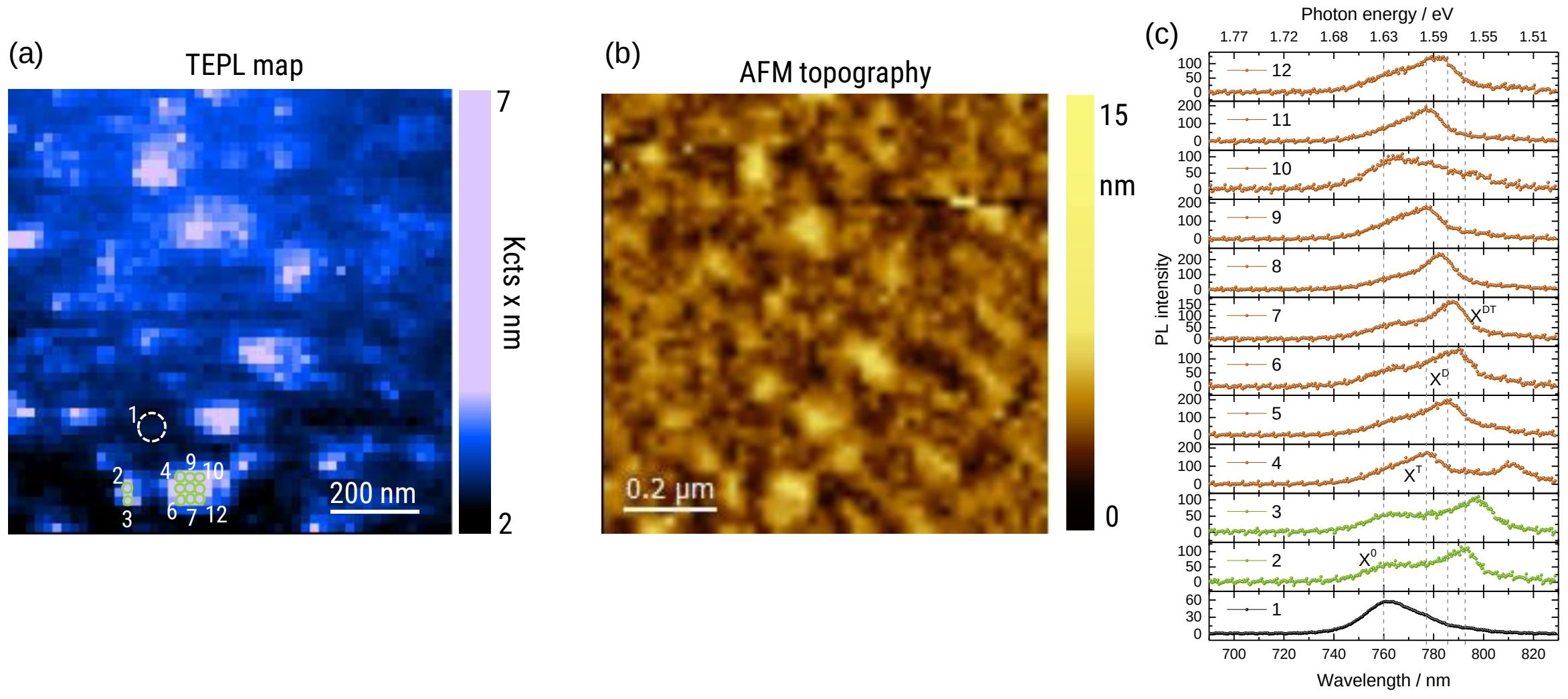

Figure SI9. TEPL map (a) and corresponding AFM topography (b) of 1L-$WSe_2$ on Au acquired simultaneously using 638 nm laser and 100 µW excitation power with a step size of 10 nm. Both TEPL map and topography show a clear correlation. The higher PL intensities stem from the nano-patches as discussed in the main text. More importantly, all the nano-patches show a clear evidence of dark excitonic emissions. For comparison PL spectra from each pixel (pixel 2 to 12) are plotted with an averaged spectra on Au (area 1) in (c). We can observe XD and XDT in TEPL spectra in agreement with Fig. 4.

## S-11: TEPL study of 1L-$WSe_2$ on Ag

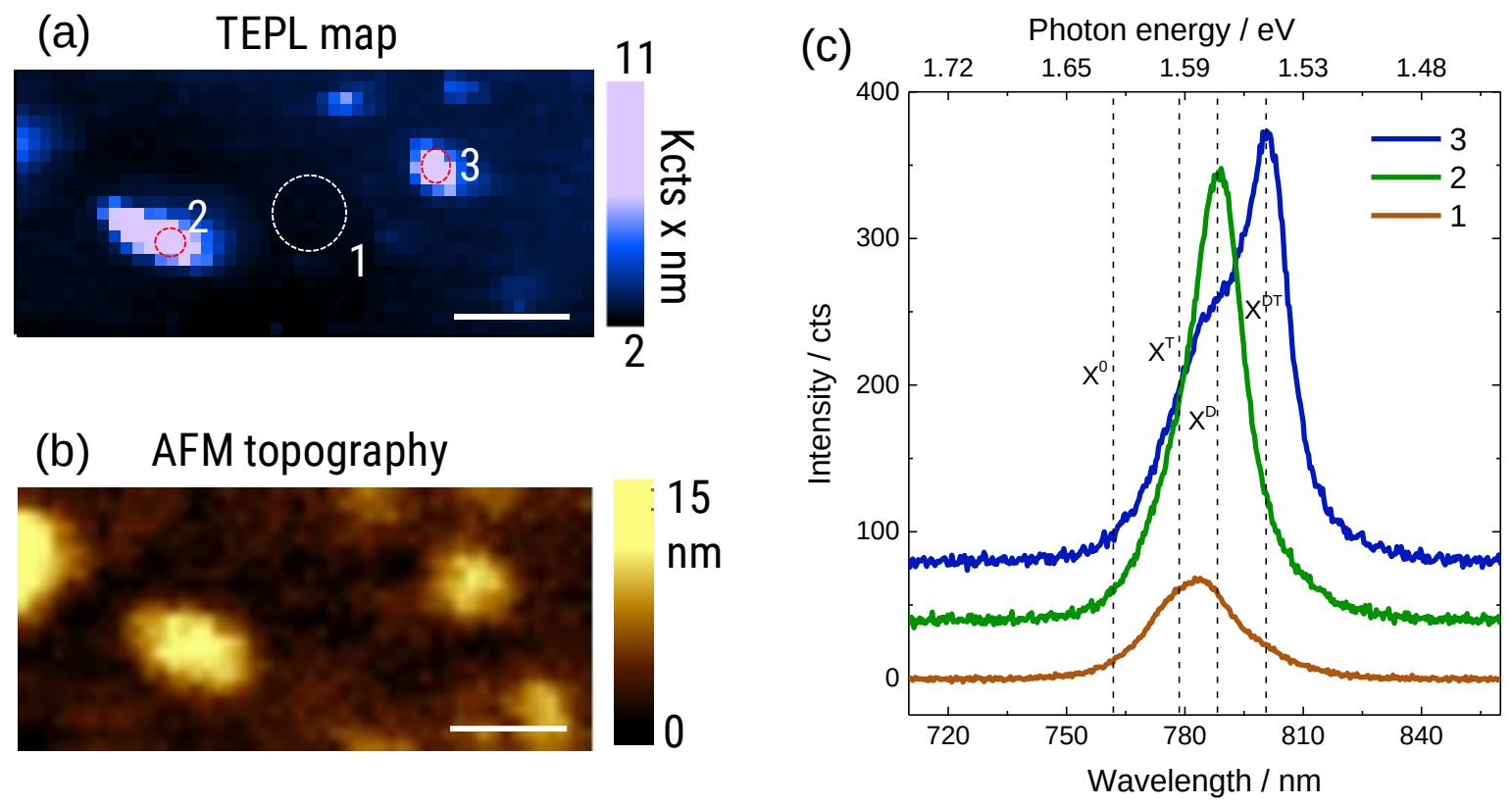

Figure SI10. TEPL map (a) and corresponding AFM topography (b) of 1L-$WSe_2$ on Ag acquired simultaneously using 638 nm laser and 100 µW excitation power with a step size of 10 nm. Similar to PL spectra of $WSe_2$ on Au we also observe dark excitonic features on Ag while scanning over the nano-patches. Scale bar is 200 nm for (a) and (b). For comparison PL spectra from three different areas on Ag are shown in (c).

## SI-12: TERS study of 1L-$WSe_2$ on Au

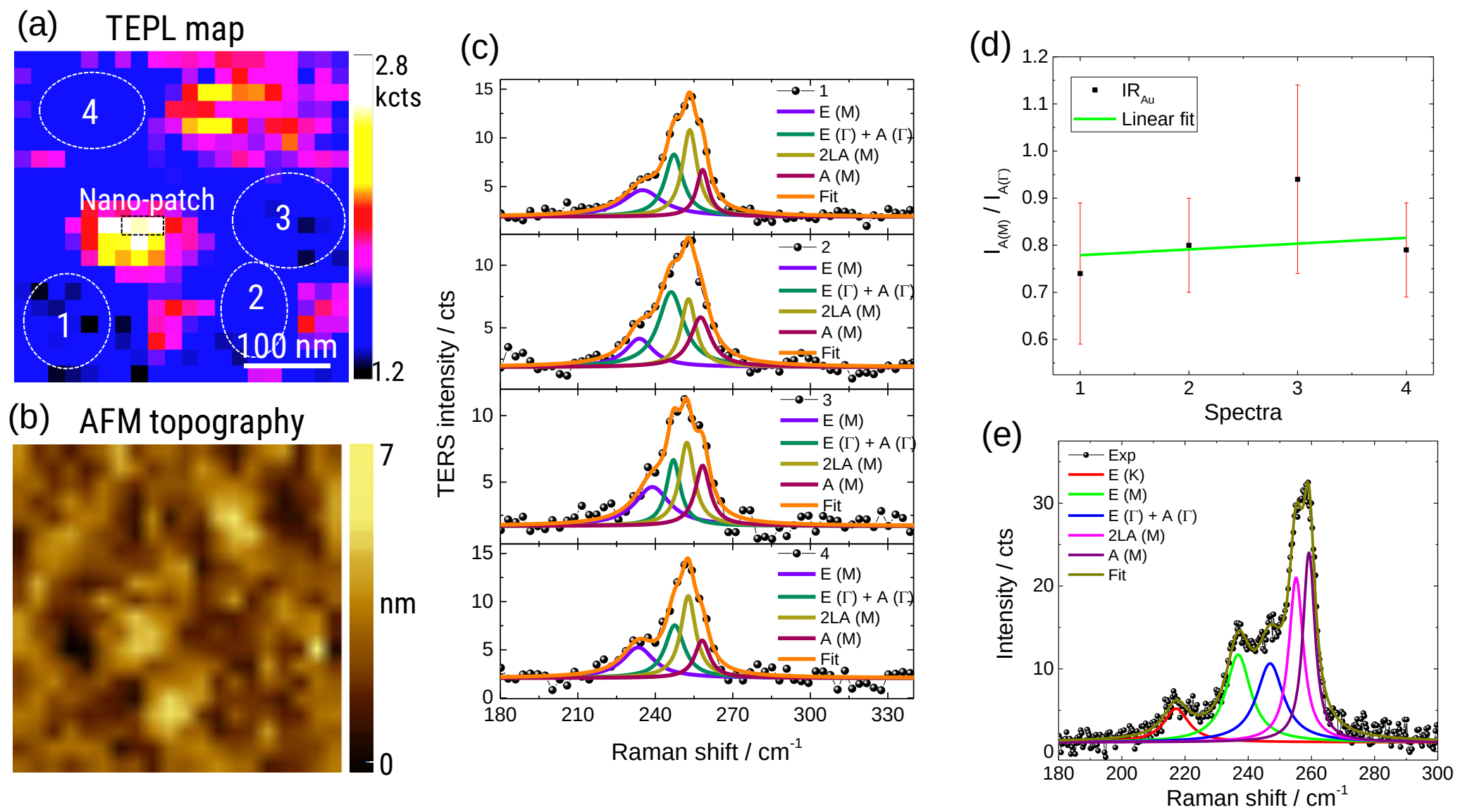

Figure SI11. TEPL map (a) and corresponding AFM topography (b) of 1L-$WSe_2$ on Au acquired simultaneously using 638 nm laser and 100 µW excitation power. The two spectra shown in Fig 4f are from area 1 and nano-patch. We also analyzed TERS on Au from other part of this map as highlighted by circles 2, 3, and 4. The spectra were fitted using Voigt functions and presented in (c). Since the appearance and intensity of the $A(M)$ mode is an indication of disorder or defects in $WSe_2$ as discussed in the main text, we also monitored the intensity ratio, $A(M)/ A(\Gamma)$ (d). An intensity ratio of 0.8 ± 0.09 on top of Au was determined. However, this value is smaller than the value determined on nano-patches (see next section). For comparison we also measured micro-Raman spectra of monolayer $WSe_2$ on Au using the same laser source and 2400 l/mm grating dispersed on EMCCD. In micro-Raman configuration we can resolve the second order phonon modes due to better spectral resolution of the measurements. Apart from two first order Raman modes E ($\Gamma$), $A$ ($\Gamma$), which are degenerate in the monolayer (see main text), we can also observe two in-plane $E$ ($K$), and $E$ ($M$) modes at the lower frequency end. At the higher frequency end the overtone of the *LA* mode at *M* point and an out-of-plane *A (M)* are detected.

## S-13: TERS study of 1L-$WSe_2$ on nano-patches

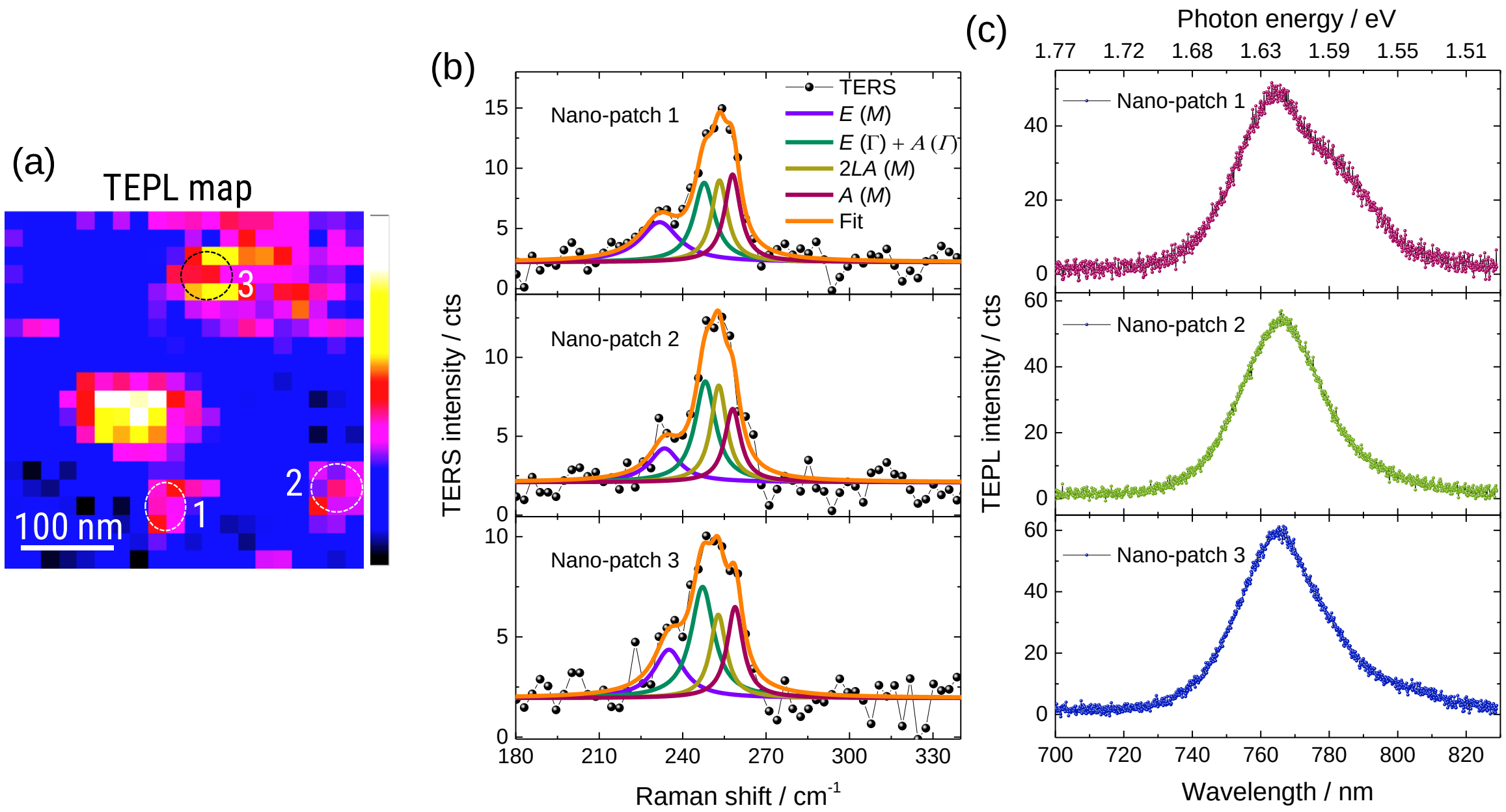

Figure SI12. TEPL map of 1L-$WSe_2$ on Au shown in Fig. S11a. For a comparative study, we analyzed TERS on all the nano-patches in this map. The corresponding TER spectra (b) and TEPL spectra (c) are presented for direct comparison. If one looks carefully, the intensity of the $A(M)$ mode in (b) is not the similar for all nano-patches. More importantly, there is a direct correlation between the intensity of $A(M)$ and the appearance of dark excitons in the PL spectra in (d). The stronger the intensity of $A(M)$, the more significant is the dark exciton emission in the PL. Since the intensity of $A(M)$ mode depends on defects (here most likely chalcogen vacancies), the density of chalcogen vacancies may be the catalyst for the radiative recombination of dark excitons in our system, which enables us to observe such excitons at room temperature.

## S-14: Local gate voltage dependent TEPL

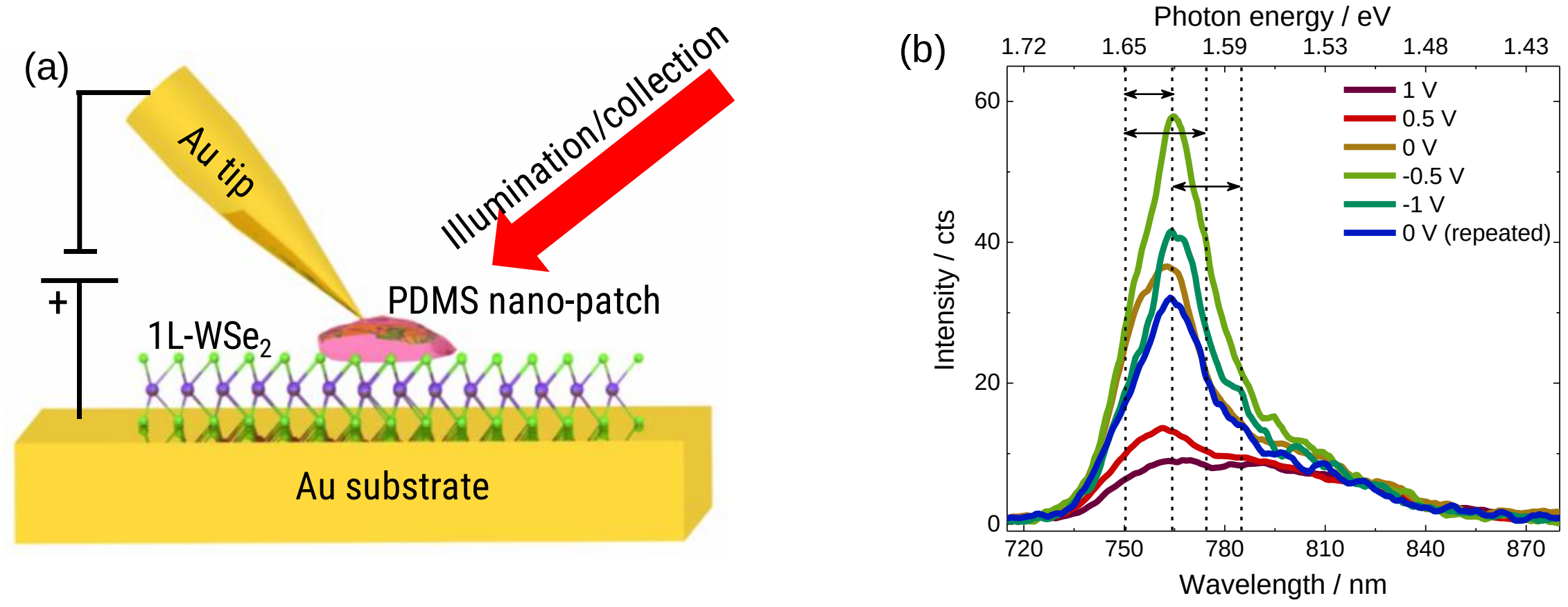

Figure SI13. Experimental configuration of the electric field dependent TEPL investigation of 1L-$WSe_2$ on Au (a). PDMS nano-patches were used as the gate dielectric and the Au tip was used as the top electrode. Typical heights of the PDMS nano-patches were (12 – 15) nm. The bias was applied to the sample. Similar to the TEPL measurements a 638 nm laser was used to excite the sample. Local gate voltage dependent TEPL spectra of 1L-$WSe_2$ (b). The measurements started from the zero bias (orange spectra) and was then gradually increased to the positive direction. After that, the negative biases were applied gradually. At the end a zero bias spectrum (blue) was acquired in the same run for comparison. In total 10 measurement runs were acquired and the same procedure was followed for all runs. The slight deviation in the two zero bias TEPL spectra may originate from the thermal drift occuring during the measurement. The drift is a common phenomenon in TEPL/TERS measurements and is unavoidable. The AFM topography measurements before and after the TEPL measurements confirm the drift in the range of few nm. As can be seen in (b), we can control the PL intensity in both directions by applying positive or negative bias. We also observed a decrease of PL intensity at higher negative bias voltage consistently. This can be due to two reasons. First, due to the thermal drift the TEPL spectra may be slightly different as shown in Fig. SI9. Secondly, the dark excitonic species may disappear at higher gate voltages as shown in recent reports.[1-3] Therefore, at a bias of -1.0 V we may observe weaker PL response compared to the bias voltage of -0.5 V. Due to the instability of the PDMS nano-patches at higher voltage under illumination we were not able to record TEPL spectra at higher bias voltages.

## S-15: AFM topography of 1L-WSe on $SiO_2$ before and after annealing

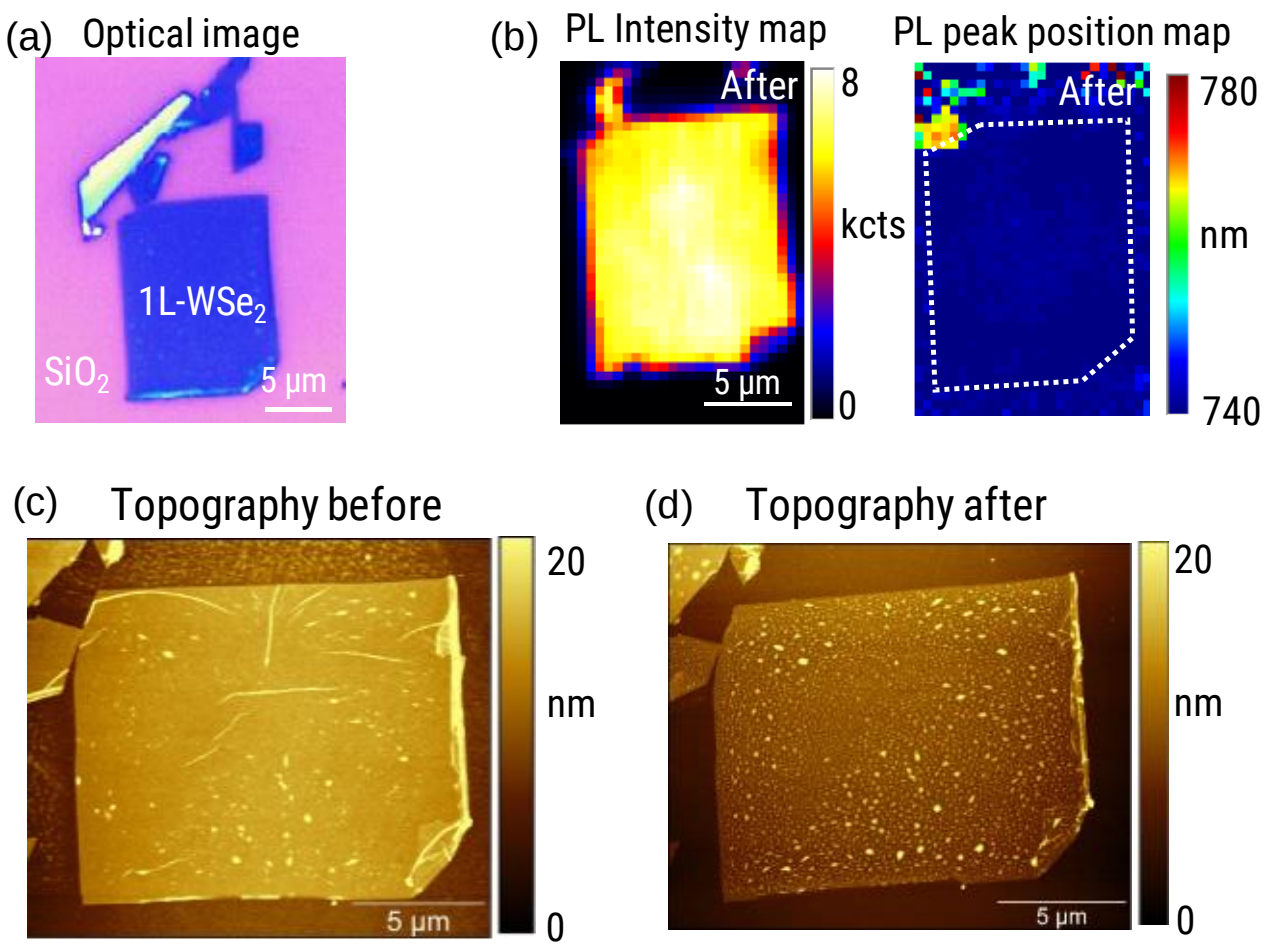

Figure SI14. Optical image of 1L-$WSe_2$ on 300 nm $SiO_2$ substrate (a). Corresponding PL intensity and peak position maps after annealing (b) as shown in Fig.6. AFM topography of the flake before (c) and after (d) annealing. Despite the presence of nano-patches after annealing, PL spectra of monolayer $WSe_2$ on $SiO_2$ show no sign of dark excitonic emission suggesting a metal substrate is required for effective electrostatic gating at room temperature (see main text).